\documentclass[traditabstract]{aa}
\usepackage{color}
\usepackage{comment}
\definecolor{AliceBlue}{rgb}{0.94,0.97,1.00}
\definecolor{AntiqueWhite1}{rgb}{1.00,0.94,0.86}
\definecolor{AntiqueWhite2}{rgb}{0.93,0.87,0.80}
\definecolor{AntiqueWhite3}{rgb}{0.80,0.75,0.69}
\definecolor{AntiqueWhite4}{rgb}{0.55,0.51,0.47}
\definecolor{AntiqueWhite}{rgb}{0.98,0.92,0.84}
\definecolor{BlanchedAlmond}{rgb}{1.00,0.92,0.80}
\definecolor{BlueViolet}{rgb}{0.54,0.17,0.89}
\definecolor{CadetBlue1}{rgb}{0.60,0.96,1.00}
\definecolor{CadetBlue2}{rgb}{0.56,0.90,0.93}
\definecolor{CadetBlue3}{rgb}{0.48,0.77,0.80}
\definecolor{CadetBlue4}{rgb}{0.33,0.53,0.55}
\definecolor{CadetBlue}{rgb}{0.37,0.62,0.63}
\definecolor{CornflowerBlue}{rgb}{0.39,0.58,0.93}
\definecolor{DarkBlue}{rgb}{0.00,0.00,0.55}
\definecolor{DarkCyan}{rgb}{0.00,0.55,0.55}
\definecolor{DarkGoldenrod1}{rgb}{1.00,0.73,0.06}
\definecolor{DarkGoldenrod2}{rgb}{0.93,0.68,0.05}
\definecolor{DarkGoldenrod3}{rgb}{0.80,0.58,0.05}
\definecolor{DarkGoldenrod4}{rgb}{0.55,0.40,0.03}
\definecolor{DarkGoldenrod}{rgb}{0.72,0.53,0.04}
\definecolor{DarkGray}{rgb}{0.66,0.66,0.66}
\definecolor{DarkGreen}{rgb}{0.00,0.39,0.00}
\definecolor{DarkGrey}{rgb}{0.66,0.66,0.66}
\definecolor{DarkKhaki}{rgb}{0.74,0.72,0.42}
\definecolor{DarkMagenta}{rgb}{0.55,0.00,0.55}
\definecolor{DarkOliveGreen1}{rgb}{0.79,1.00,0.44}
\definecolor{DarkOliveGreen2}{rgb}{0.74,0.93,0.41}
\definecolor{DarkOliveGreen3}{rgb}{0.64,0.80,0.35}
\definecolor{DarkOliveGreen4}{rgb}{0.43,0.55,0.24}
\definecolor{DarkOliveGreen}{rgb}{0.33,0.42,0.18}
\definecolor{DarkOrange1}{rgb}{1.00,0.50,0.00}
\definecolor{DarkOrange2}{rgb}{0.93,0.46,0.00}
\definecolor{DarkOrange3}{rgb}{0.80,0.40,0.00}
\definecolor{DarkOrange4}{rgb}{0.55,0.27,0.00}
\definecolor{DarkOrange}{rgb}{1.00,0.55,0.00}
\definecolor{DarkOrchid1}{rgb}{0.75,0.24,1.00}
\definecolor{DarkOrchid2}{rgb}{0.70,0.23,0.93}
\definecolor{DarkOrchid3}{rgb}{0.60,0.20,0.80}
\definecolor{DarkOrchid4}{rgb}{0.41,0.13,0.55}
\definecolor{DarkOrchid}{rgb}{0.60,0.20,0.80}
\definecolor{DarkRed}{rgb}{0.55,0.00,0.00}
\definecolor{DarkSalmon}{rgb}{0.91,0.59,0.48}
\definecolor{DarkSeaGreen1}{rgb}{0.76,1.00,0.76}
\definecolor{DarkSeaGreen2}{rgb}{0.71,0.93,0.71}
\definecolor{DarkSeaGreen3}{rgb}{0.61,0.80,0.61}
\definecolor{DarkSeaGreen4}{rgb}{0.41,0.55,0.41}
\definecolor{DarkSeaGreen}{rgb}{0.56,0.74,0.56}
\definecolor{DarkSlateBlue}{rgb}{0.28,0.24,0.55}
\definecolor{DarkSlateGray1}{rgb}{0.59,1.00,1.00}
\definecolor{DarkSlateGray2}{rgb}{0.55,0.93,0.93}
\definecolor{DarkSlateGray3}{rgb}{0.47,0.80,0.80}
\definecolor{DarkSlateGray4}{rgb}{0.32,0.55,0.55}
\definecolor{DarkSlateGray}{rgb}{0.18,0.31,0.31}
\definecolor{DarkSlateGrey}{rgb}{0.18,0.31,0.31}
\definecolor{DarkTurquoise}{rgb}{0.00,0.81,0.82}
\definecolor{DarkViolet}{rgb}{0.58,0.00,0.83}
\definecolor{DeepPink1}{rgb}{1.00,0.08,0.58}
\definecolor{DeepPink2}{rgb}{0.93,0.07,0.54}
\definecolor{DeepPink3}{rgb}{0.80,0.06,0.46}
\definecolor{DeepPink4}{rgb}{0.55,0.04,0.31}
\definecolor{DeepPink}{rgb}{1.00,0.08,0.58}
\definecolor{DeepSkyBlue1}{rgb}{0.00,0.75,1.00}
\definecolor{DeepSkyBlue2}{rgb}{0.00,0.70,0.93}
\definecolor{DeepSkyBlue3}{rgb}{0.00,0.60,0.80}
\definecolor{DeepSkyBlue4}{rgb}{0.00,0.41,0.55}
\definecolor{DeepSkyBlue}{rgb}{0.00,0.75,1.00}
\definecolor{DimGray}{rgb}{0.41,0.41,0.41}
\definecolor{DimGrey}{rgb}{0.41,0.41,0.41}
\definecolor{DodgerBlue1}{rgb}{0.12,0.56,1.00}
\definecolor{DodgerBlue2}{rgb}{0.11,0.53,0.93}
\definecolor{DodgerBlue3}{rgb}{0.09,0.45,0.80}
\definecolor{DodgerBlue4}{rgb}{0.06,0.31,0.55}
\definecolor{DodgerBlue}{rgb}{0.12,0.56,1.00}
\definecolor{FloralWhite}{rgb}{1.00,0.98,0.94}
\definecolor{ForestGreen}{rgb}{0.13,0.55,0.13}
\definecolor{GhostWhite}{rgb}{0.97,0.97,1.00}
\definecolor{GreenYellow}{rgb}{0.68,1.00,0.18}
\definecolor{HotPink1}{rgb}{1.00,0.43,0.71}
\definecolor{HotPink2}{rgb}{0.93,0.42,0.65}
\definecolor{HotPink3}{rgb}{0.80,0.38,0.56}
\definecolor{HotPink4}{rgb}{0.55,0.23,0.38}
\definecolor{HotPink}{rgb}{1.00,0.41,0.71}
\definecolor{IndianRed1}{rgb}{1.00,0.42,0.42}
\definecolor{IndianRed2}{rgb}{0.93,0.39,0.39}
\definecolor{IndianRed3}{rgb}{0.80,0.33,0.33}
\definecolor{IndianRed4}{rgb}{0.55,0.23,0.23}
\definecolor{IndianRed}{rgb}{0.80,0.36,0.36}
\definecolor{LavenderBlush1}{rgb}{1.00,0.94,0.96}
\definecolor{LavenderBlush2}{rgb}{0.93,0.88,0.90}
\definecolor{LavenderBlush3}{rgb}{0.80,0.76,0.77}
\definecolor{LavenderBlush4}{rgb}{0.55,0.51,0.53}
\definecolor{LavenderBlush}{rgb}{1.00,0.94,0.96}
\definecolor{LawnGreen}{rgb}{0.49,0.99,0.00}
\definecolor{LemonChiffon1}{rgb}{1.00,0.98,0.80}
\definecolor{LemonChiffon2}{rgb}{0.93,0.91,0.75}
\definecolor{LemonChiffon3}{rgb}{0.80,0.79,0.65}
\definecolor{LemonChiffon4}{rgb}{0.55,0.54,0.44}
\definecolor{LemonChiffon}{rgb}{1.00,0.98,0.80}
\definecolor{LightBlue1}{rgb}{0.75,0.94,1.00}
\definecolor{LightBlue2}{rgb}{0.70,0.87,0.93}
\definecolor{LightBlue3}{rgb}{0.60,0.75,0.80}
\definecolor{LightBlue4}{rgb}{0.41,0.51,0.55}
\definecolor{LightBlue}{rgb}{0.68,0.85,0.90}
\definecolor{LightCoral}{rgb}{0.94,0.50,0.50}
\definecolor{LightCyan1}{rgb}{0.88,1.00,1.00}
\definecolor{LightCyan2}{rgb}{0.82,0.93,0.93}
\definecolor{LightCyan3}{rgb}{0.71,0.80,0.80}
\definecolor{LightCyan4}{rgb}{0.48,0.55,0.55}
\definecolor{LightCyan}{rgb}{0.88,1.00,1.00}
\definecolor{LightGoldenrod1}{rgb}{1.00,0.93,0.55}
\definecolor{LightGoldenrod2}{rgb}{0.93,0.86,0.51}
\definecolor{LightGoldenrod3}{rgb}{0.80,0.75,0.44}
\definecolor{LightGoldenrod4}{rgb}{0.55,0.51,0.30}
\definecolor{LightGoldenrodYellow}{rgb}{0.98,0.98,0.82}
\definecolor{LightGoldenrod}{rgb}{0.93,0.87,0.51}
\definecolor{LightGray}{rgb}{0.83,0.83,0.83}
\definecolor{LightGreen}{rgb}{0.56,0.93,0.56}
\definecolor{LightGrey}{rgb}{0.83,0.83,0.83}
\definecolor{LightPink1}{rgb}{1.00,0.68,0.73}
\definecolor{LightPink2}{rgb}{0.93,0.64,0.68}
\definecolor{LightPink3}{rgb}{0.80,0.55,0.58}
\definecolor{LightPink4}{rgb}{0.55,0.37,0.40}
\definecolor{LightPink}{rgb}{1.00,0.71,0.76}
\definecolor{LightSalmon1}{rgb}{1.00,0.63,0.48}
\definecolor{LightSalmon2}{rgb}{0.93,0.58,0.45}
\definecolor{LightSalmon3}{rgb}{0.80,0.51,0.38}
\definecolor{LightSalmon4}{rgb}{0.55,0.34,0.26}
\definecolor{LightSalmon}{rgb}{1.00,0.63,0.48}
\definecolor{LightSeaGreen}{rgb}{0.13,0.70,0.67}
\definecolor{LightSkyBlue1}{rgb}{0.69,0.89,1.00}
\definecolor{LightSkyBlue2}{rgb}{0.64,0.83,0.93}
\definecolor{LightSkyBlue3}{rgb}{0.55,0.71,0.80}
\definecolor{LightSkyBlue4}{rgb}{0.38,0.48,0.55}
\definecolor{LightSkyBlue}{rgb}{0.53,0.81,0.98}
\definecolor{LightSlateBlue}{rgb}{0.52,0.44,1.00}
\definecolor{LightSlateGray}{rgb}{0.47,0.53,0.60}
\definecolor{LightSlateGrey}{rgb}{0.47,0.53,0.60}
\definecolor{LightSteelBlue1}{rgb}{0.79,0.88,1.00}
\definecolor{LightSteelBlue2}{rgb}{0.74,0.82,0.93}
\definecolor{LightSteelBlue3}{rgb}{0.64,0.71,0.80}
\definecolor{LightSteelBlue4}{rgb}{0.43,0.48,0.55}
\definecolor{LightSteelBlue}{rgb}{0.69,0.77,0.87}
\definecolor{LightYellow1}{rgb}{1.00,1.00,0.88}
\definecolor{LightYellow2}{rgb}{0.93,0.93,0.82}
\definecolor{LightYellow3}{rgb}{0.80,0.80,0.71}
\definecolor{LightYellow4}{rgb}{0.55,0.55,0.48}
\definecolor{LightYellow}{rgb}{1.00,1.00,0.88}
\definecolor{LimeGreen}{rgb}{0.20,0.80,0.20}
\definecolor{MediumAquamarine}{rgb}{0.40,0.80,0.67}
\definecolor{MediumBlue}{rgb}{0.00,0.00,0.80}
\definecolor{MediumOrchid1}{rgb}{0.88,0.40,1.00}
\definecolor{MediumOrchid2}{rgb}{0.82,0.37,0.93}
\definecolor{MediumOrchid3}{rgb}{0.71,0.32,0.80}
\definecolor{MediumOrchid4}{rgb}{0.48,0.22,0.55}
\definecolor{MediumOrchid}{rgb}{0.73,0.33,0.83}
\definecolor{MediumPurple1}{rgb}{0.67,0.51,1.00}
\definecolor{MediumPurple2}{rgb}{0.62,0.47,0.93}
\definecolor{MediumPurple3}{rgb}{0.54,0.41,0.80}
\definecolor{MediumPurple4}{rgb}{0.36,0.28,0.55}
\definecolor{MediumPurple}{rgb}{0.58,0.44,0.86}
\definecolor{MediumSeaGreen}{rgb}{0.24,0.70,0.44}
\definecolor{MediumSlateBlue}{rgb}{0.48,0.41,0.93}
\definecolor{MediumSpringGreen}{rgb}{0.00,0.98,0.60}
\definecolor{MediumTurquoise}{rgb}{0.28,0.82,0.80}
\definecolor{MediumVioletRed}{rgb}{0.78,0.08,0.52}
\definecolor{MidnightBlue}{rgb}{0.10,0.10,0.44}
\definecolor{MintCream}{rgb}{0.96,1.00,0.98}
\definecolor{MistyRose1}{rgb}{1.00,0.89,0.88}
\definecolor{MistyRose2}{rgb}{0.93,0.84,0.82}
\definecolor{MistyRose3}{rgb}{0.80,0.72,0.71}
\definecolor{MistyRose4}{rgb}{0.55,0.49,0.48}
\definecolor{MistyRose}{rgb}{1.00,0.89,0.88}
\definecolor{NavajoWhite1}{rgb}{1.00,0.87,0.68}
\definecolor{NavajoWhite2}{rgb}{0.93,0.81,0.63}
\definecolor{NavajoWhite3}{rgb}{0.80,0.70,0.55}
\definecolor{NavajoWhite4}{rgb}{0.55,0.47,0.37}
\definecolor{NavajoWhite}{rgb}{1.00,0.87,0.68}
\definecolor{NavyBlue}{rgb}{0.00,0.00,0.50}
\definecolor{OldLace}{rgb}{0.99,0.96,0.90}
\definecolor{OliveDrab1}{rgb}{0.75,1.00,0.24}
\definecolor{OliveDrab2}{rgb}{0.70,0.93,0.23}
\definecolor{OliveDrab3}{rgb}{0.60,0.80,0.20}
\definecolor{OliveDrab4}{rgb}{0.41,0.55,0.13}
\definecolor{OliveDrab}{rgb}{0.42,0.56,0.14}
\definecolor{OrangeRed1}{rgb}{1.00,0.27,0.00}
\definecolor{OrangeRed2}{rgb}{0.93,0.25,0.00}
\definecolor{OrangeRed3}{rgb}{0.80,0.22,0.00}
\definecolor{OrangeRed4}{rgb}{0.55,0.15,0.00}
\definecolor{OrangeRed}{rgb}{1.00,0.27,0.00}
\definecolor{PaleGoldenrod}{rgb}{0.93,0.91,0.67}
\definecolor{PaleGreen1}{rgb}{0.60,1.00,0.60}
\definecolor{PaleGreen2}{rgb}{0.56,0.93,0.56}
\definecolor{PaleGreen3}{rgb}{0.49,0.80,0.49}
\definecolor{PaleGreen4}{rgb}{0.33,0.55,0.33}
\definecolor{PaleGreen}{rgb}{0.60,0.98,0.60}
\definecolor{PaleTurquoise1}{rgb}{0.73,1.00,1.00}
\definecolor{PaleTurquoise2}{rgb}{0.68,0.93,0.93}
\definecolor{PaleTurquoise3}{rgb}{0.59,0.80,0.80}
\definecolor{PaleTurquoise4}{rgb}{0.40,0.55,0.55}
\definecolor{PaleTurquoise}{rgb}{0.69,0.93,0.93}
\definecolor{PaleVioletRed1}{rgb}{1.00,0.51,0.67}
\definecolor{PaleVioletRed2}{rgb}{0.93,0.47,0.62}
\definecolor{PaleVioletRed3}{rgb}{0.80,0.41,0.54}
\definecolor{PaleVioletRed4}{rgb}{0.55,0.28,0.36}
\definecolor{PaleVioletRed}{rgb}{0.86,0.44,0.58}
\definecolor{PapayaWhip}{rgb}{1.00,0.94,0.84}
\definecolor{PeachPuff1}{rgb}{1.00,0.85,0.73}
\definecolor{PeachPuff2}{rgb}{0.93,0.80,0.68}
\definecolor{PeachPuff3}{rgb}{0.80,0.69,0.58}
\definecolor{PeachPuff4}{rgb}{0.55,0.47,0.40}
\definecolor{PeachPuff}{rgb}{1.00,0.85,0.73}
\definecolor{PowderBlue}{rgb}{0.69,0.88,0.90}
\definecolor{RosyBrown1}{rgb}{1.00,0.76,0.76}
\definecolor{RosyBrown2}{rgb}{0.93,0.71,0.71}
\definecolor{RosyBrown3}{rgb}{0.80,0.61,0.61}
\definecolor{RosyBrown4}{rgb}{0.55,0.41,0.41}
\definecolor{RosyBrown}{rgb}{0.74,0.56,0.56}
\definecolor{RoyalBlue1}{rgb}{0.28,0.46,1.00}
\definecolor{RoyalBlue2}{rgb}{0.26,0.43,0.93}
\definecolor{RoyalBlue3}{rgb}{0.23,0.37,0.80}
\definecolor{RoyalBlue4}{rgb}{0.15,0.25,0.55}
\definecolor{RoyalBlue}{rgb}{0.25,0.41,0.88}
\definecolor{SaddleBrown}{rgb}{0.55,0.27,0.07}
\definecolor{SandyBrown}{rgb}{0.96,0.64,0.38}
\definecolor{SeaGreen1}{rgb}{0.33,1.00,0.62}
\definecolor{SeaGreen2}{rgb}{0.31,0.93,0.58}
\definecolor{SeaGreen3}{rgb}{0.26,0.80,0.50}
\definecolor{SeaGreen4}{rgb}{0.18,0.55,0.34}
\definecolor{SeaGreen}{rgb}{0.18,0.55,0.34}
\definecolor{SkyBlue1}{rgb}{0.53,0.81,1.00}
\definecolor{SkyBlue2}{rgb}{0.49,0.75,0.93}
\definecolor{SkyBlue3}{rgb}{0.42,0.65,0.80}
\definecolor{SkyBlue4}{rgb}{0.29,0.44,0.55}
\definecolor{SkyBlue}{rgb}{0.53,0.81,0.92}
\definecolor{SlateBlue1}{rgb}{0.51,0.44,1.00}
\definecolor{SlateBlue2}{rgb}{0.48,0.40,0.93}
\definecolor{SlateBlue3}{rgb}{0.41,0.35,0.80}
\definecolor{SlateBlue4}{rgb}{0.28,0.24,0.55}
\definecolor{SlateBlue}{rgb}{0.42,0.35,0.80}
\definecolor{SlateGray1}{rgb}{0.78,0.89,1.00}
\definecolor{SlateGray2}{rgb}{0.73,0.83,0.93}
\definecolor{SlateGray3}{rgb}{0.62,0.71,0.80}
\definecolor{SlateGray4}{rgb}{0.42,0.48,0.55}
\definecolor{SlateGray}{rgb}{0.44,0.50,0.56}
\definecolor{SlateGrey}{rgb}{0.44,0.50,0.56}
\definecolor{SpringGreen1}{rgb}{0.00,1.00,0.50}
\definecolor{SpringGreen2}{rgb}{0.00,0.93,0.46}
\definecolor{SpringGreen3}{rgb}{0.00,0.80,0.40}
\definecolor{SpringGreen4}{rgb}{0.00,0.55,0.27}
\definecolor{SpringGreen}{rgb}{0.00,1.00,0.50}
\definecolor{SteelBlue1}{rgb}{0.39,0.72,1.00}
\definecolor{SteelBlue2}{rgb}{0.36,0.67,0.93}
\definecolor{SteelBlue3}{rgb}{0.31,0.58,0.80}
\definecolor{SteelBlue4}{rgb}{0.21,0.39,0.55}
\definecolor{SteelBlue}{rgb}{0.27,0.51,0.71}
\definecolor{VioletRed1}{rgb}{1.00,0.24,0.59}
\definecolor{VioletRed2}{rgb}{0.93,0.23,0.55}
\definecolor{VioletRed3}{rgb}{0.80,0.20,0.47}
\definecolor{VioletRed4}{rgb}{0.55,0.13,0.32}
\definecolor{VioletRed}{rgb}{0.82,0.13,0.56}
\definecolor{WhiteSmoke}{rgb}{0.96,0.96,0.96}
\definecolor{YellowGreen}{rgb}{0.60,0.80,0.20}
\definecolor{aliceblue}{rgb}{0.94,0.97,1.00}
\definecolor{antiquewhite}{rgb}{0.98,0.92,0.84}
\definecolor{aquamarine1}{rgb}{0.50,1.00,0.83}
\definecolor{aquamarine2}{rgb}{0.46,0.93,0.78}
\definecolor{aquamarine3}{rgb}{0.40,0.80,0.67}
\definecolor{aquamarine4}{rgb}{0.27,0.55,0.45}
\definecolor{aquamarine}{rgb}{0.50,1.00,0.83}
\definecolor{azure1}{rgb}{0.94,1.00,1.00}
\definecolor{azure2}{rgb}{0.88,0.93,0.93}
\definecolor{azure3}{rgb}{0.76,0.80,0.80}
\definecolor{azure4}{rgb}{0.51,0.55,0.55}
\definecolor{azure}{rgb}{0.94,1.00,1.00}
\definecolor{beige}{rgb}{0.96,0.96,0.86}
\definecolor{bisque1}{rgb}{1.00,0.89,0.77}
\definecolor{bisque2}{rgb}{0.93,0.84,0.72}
\definecolor{bisque3}{rgb}{0.80,0.72,0.62}
\definecolor{bisque4}{rgb}{0.55,0.49,0.42}
\definecolor{bisque}{rgb}{1.00,0.89,0.77}
\definecolor{black}{rgb}{0.00,0.00,0.00}
\definecolor{blanchedalmond}{rgb}{1.00,0.92,0.80}
\definecolor{blue1}{rgb}{0.00,0.00,1.00}
\definecolor{blue2}{rgb}{0.00,0.00,0.93}
\definecolor{blue3}{rgb}{0.00,0.00,0.80}
\definecolor{blue4}{rgb}{0.00,0.00,0.55}
\definecolor{blueviolet}{rgb}{0.54,0.17,0.89}
\definecolor{blue}{rgb}{0.00,0.00,1.00}
\definecolor{brown1}{rgb}{1.00,0.25,0.25}
\definecolor{brown2}{rgb}{0.93,0.23,0.23}
\definecolor{brown3}{rgb}{0.80,0.20,0.20}
\definecolor{brown4}{rgb}{0.55,0.14,0.14}
\definecolor{brown}{rgb}{0.65,0.16,0.16}
\definecolor{burlywood1}{rgb}{1.00,0.83,0.61}
\definecolor{burlywood2}{rgb}{0.93,0.77,0.57}
\definecolor{burlywood3}{rgb}{0.80,0.67,0.49}
\definecolor{burlywood4}{rgb}{0.55,0.45,0.33}
\definecolor{burlywood}{rgb}{0.87,0.72,0.53}
\definecolor{cadetblue}{rgb}{0.37,0.62,0.63}
\definecolor{chartreuse1}{rgb}{0.50,1.00,0.00}
\definecolor{chartreuse2}{rgb}{0.46,0.93,0.00}
\definecolor{chartreuse3}{rgb}{0.40,0.80,0.00}
\definecolor{chartreuse4}{rgb}{0.27,0.55,0.00}
\definecolor{chartreuse}{rgb}{0.50,1.00,0.00}
\definecolor{chocolate1}{rgb}{1.00,0.50,0.14}
\definecolor{chocolate2}{rgb}{0.93,0.46,0.13}
\definecolor{chocolate3}{rgb}{0.80,0.40,0.11}
\definecolor{chocolate4}{rgb}{0.55,0.27,0.07}
\definecolor{chocolate}{rgb}{0.82,0.41,0.12}
\definecolor{coral1}{rgb}{1.00,0.45,0.34}
\definecolor{coral2}{rgb}{0.93,0.42,0.31}
\definecolor{coral3}{rgb}{0.80,0.36,0.27}
\definecolor{coral4}{rgb}{0.55,0.24,0.18}
\definecolor{coral}{rgb}{1.00,0.50,0.31}
\definecolor{cornflowerblue}{rgb}{0.39,0.58,0.93}
\definecolor{cornsilk1}{rgb}{1.00,0.97,0.86}
\definecolor{cornsilk2}{rgb}{0.93,0.91,0.80}
\definecolor{cornsilk3}{rgb}{0.80,0.78,0.69}
\definecolor{cornsilk4}{rgb}{0.55,0.53,0.47}
\definecolor{cornsilk}{rgb}{1.00,0.97,0.86}
\definecolor{cyan1}{rgb}{0.00,1.00,1.00}
\definecolor{cyan2}{rgb}{0.00,0.93,0.93}
\definecolor{cyan3}{rgb}{0.00,0.80,0.80}
\definecolor{cyan4}{rgb}{0.00,0.55,0.55}
\definecolor{cyan}{rgb}{0.00,1.00,1.00}
\definecolor{darkblue}{rgb}{0.00,0.00,0.55}
\definecolor{darkcyan}{rgb}{0.00,0.55,0.55}
\definecolor{darkgoldenrod}{rgb}{0.72,0.53,0.04}
\definecolor{darkgray}{rgb}{0.66,0.66,0.66}
\definecolor{darkgreen}{rgb}{0.00,0.39,0.00}
\definecolor{darkgrey}{rgb}{0.66,0.66,0.66}
\definecolor{darkkhaki}{rgb}{0.74,0.72,0.42}
\definecolor{darkmagenta}{rgb}{0.55,0.00,0.55}
\definecolor{darkolive}{rgb}{0.33,0.42,0.18}
\definecolor{darkorange}{rgb}{1.00,0.55,0.00}
\definecolor{darkorchid}{rgb}{0.60,0.20,0.80}
\definecolor{darkred}{rgb}{0.55,0.00,0.00}
\definecolor{darksalmon}{rgb}{0.91,0.59,0.48}
\definecolor{darksea}{rgb}{0.56,0.74,0.56}
\definecolor{darkslate}{rgb}{0.18,0.31,0.31}
\definecolor{darkslate}{rgb}{0.18,0.31,0.31}
\definecolor{darkslate}{rgb}{0.28,0.24,0.55}
\definecolor{darkturquoise}{rgb}{0.00,0.81,0.82}
\definecolor{darkviolet}{rgb}{0.58,0.00,0.83}
\definecolor{deeppink}{rgb}{1.00,0.08,0.58}
\definecolor{deepsky}{rgb}{0.00,0.75,1.00}
\definecolor{dimgray}{rgb}{0.41,0.41,0.41}
\definecolor{dimgrey}{rgb}{0.41,0.41,0.41}
\definecolor{dodgerblue}{rgb}{0.12,0.56,1.00}
\definecolor{firebrick1}{rgb}{1.00,0.19,0.19}
\definecolor{firebrick2}{rgb}{0.93,0.17,0.17}
\definecolor{firebrick3}{rgb}{0.80,0.15,0.15}
\definecolor{firebrick4}{rgb}{0.55,0.10,0.10}
\definecolor{firebrick}{rgb}{0.70,0.13,0.13}
\definecolor{floralwhite}{rgb}{1.00,0.98,0.94}
\definecolor{forestgreen}{rgb}{0.13,0.55,0.13}
\definecolor{gainsboro}{rgb}{0.86,0.86,0.86}
\definecolor{ghostwhite}{rgb}{0.97,0.97,1.00}
\definecolor{gold1}{rgb}{1.00,0.84,0.00}
\definecolor{gold2}{rgb}{0.93,0.79,0.00}
\definecolor{gold3}{rgb}{0.80,0.68,0.00}
\definecolor{gold4}{rgb}{0.55,0.46,0.00}
\definecolor{goldenrod1}{rgb}{1.00,0.76,0.15}
\definecolor{goldenrod2}{rgb}{0.93,0.71,0.13}
\definecolor{goldenrod3}{rgb}{0.80,0.61,0.11}
\definecolor{goldenrod4}{rgb}{0.55,0.41,0.08}
\definecolor{goldenrod}{rgb}{0.85,0.65,0.13}
\definecolor{gold}{rgb}{1.00,0.84,0.00}
\definecolor{gray0}{rgb}{0.00,0.00,0.00}
\definecolor{gray100}{rgb}{1.00,1.00,1.00}
\definecolor{gray10}{rgb}{0.10,0.10,0.10}
\definecolor{gray11}{rgb}{0.11,0.11,0.11}
\definecolor{gray12}{rgb}{0.12,0.12,0.12}
\definecolor{gray13}{rgb}{0.13,0.13,0.13}
\definecolor{gray14}{rgb}{0.14,0.14,0.14}
\definecolor{gray15}{rgb}{0.15,0.15,0.15}
\definecolor{gray16}{rgb}{0.16,0.16,0.16}
\definecolor{gray17}{rgb}{0.17,0.17,0.17}
\definecolor{gray18}{rgb}{0.18,0.18,0.18}
\definecolor{gray19}{rgb}{0.19,0.19,0.19}
\definecolor{gray1}{rgb}{0.01,0.01,0.01}
\definecolor{gray20}{rgb}{0.20,0.20,0.20}
\definecolor{gray21}{rgb}{0.21,0.21,0.21}
\definecolor{gray22}{rgb}{0.22,0.22,0.22}
\definecolor{gray23}{rgb}{0.23,0.23,0.23}
\definecolor{gray24}{rgb}{0.24,0.24,0.24}
\definecolor{gray25}{rgb}{0.25,0.25,0.25}
\definecolor{gray26}{rgb}{0.26,0.26,0.26}
\definecolor{gray27}{rgb}{0.27,0.27,0.27}
\definecolor{gray28}{rgb}{0.28,0.28,0.28}
\definecolor{gray29}{rgb}{0.29,0.29,0.29}
\definecolor{gray2}{rgb}{0.02,0.02,0.02}
\definecolor{gray30}{rgb}{0.30,0.30,0.30}
\definecolor{gray31}{rgb}{0.31,0.31,0.31}
\definecolor{gray32}{rgb}{0.32,0.32,0.32}
\definecolor{gray33}{rgb}{0.33,0.33,0.33}
\definecolor{gray34}{rgb}{0.34,0.34,0.34}
\definecolor{gray35}{rgb}{0.35,0.35,0.35}
\definecolor{gray36}{rgb}{0.36,0.36,0.36}
\definecolor{gray37}{rgb}{0.37,0.37,0.37}
\definecolor{gray38}{rgb}{0.38,0.38,0.38}
\definecolor{gray39}{rgb}{0.39,0.39,0.39}
\definecolor{gray3}{rgb}{0.03,0.03,0.03}
\definecolor{gray40}{rgb}{0.40,0.40,0.40}
\definecolor{gray41}{rgb}{0.41,0.41,0.41}
\definecolor{gray42}{rgb}{0.42,0.42,0.42}
\definecolor{gray43}{rgb}{0.43,0.43,0.43}
\definecolor{gray44}{rgb}{0.44,0.44,0.44}
\definecolor{gray45}{rgb}{0.45,0.45,0.45}
\definecolor{gray46}{rgb}{0.46,0.46,0.46}
\definecolor{gray47}{rgb}{0.47,0.47,0.47}
\definecolor{gray48}{rgb}{0.48,0.48,0.48}
\definecolor{gray49}{rgb}{0.49,0.49,0.49}
\definecolor{gray4}{rgb}{0.04,0.04,0.04}
\definecolor{gray50}{rgb}{0.50,0.50,0.50}
\definecolor{gray51}{rgb}{0.51,0.51,0.51}
\definecolor{gray52}{rgb}{0.52,0.52,0.52}
\definecolor{gray53}{rgb}{0.53,0.53,0.53}
\definecolor{gray54}{rgb}{0.54,0.54,0.54}
\definecolor{gray55}{rgb}{0.55,0.55,0.55}
\definecolor{gray56}{rgb}{0.56,0.56,0.56}
\definecolor{gray57}{rgb}{0.57,0.57,0.57}
\definecolor{gray58}{rgb}{0.58,0.58,0.58}
\definecolor{gray59}{rgb}{0.59,0.59,0.59}
\definecolor{gray5}{rgb}{0.05,0.05,0.05}
\definecolor{gray60}{rgb}{0.60,0.60,0.60}
\definecolor{gray61}{rgb}{0.61,0.61,0.61}
\definecolor{gray62}{rgb}{0.62,0.62,0.62}
\definecolor{gray63}{rgb}{0.63,0.63,0.63}
\definecolor{gray64}{rgb}{0.64,0.64,0.64}
\definecolor{gray65}{rgb}{0.65,0.65,0.65}
\definecolor{gray66}{rgb}{0.66,0.66,0.66}
\definecolor{gray67}{rgb}{0.67,0.67,0.67}
\definecolor{gray68}{rgb}{0.68,0.68,0.68}
\definecolor{gray69}{rgb}{0.69,0.69,0.69}
\definecolor{gray6}{rgb}{0.06,0.06,0.06}
\definecolor{gray70}{rgb}{0.70,0.70,0.70}
\definecolor{gray71}{rgb}{0.71,0.71,0.71}
\definecolor{gray72}{rgb}{0.72,0.72,0.72}
\definecolor{gray73}{rgb}{0.73,0.73,0.73}
\definecolor{gray74}{rgb}{0.74,0.74,0.74}
\definecolor{gray75}{rgb}{0.75,0.75,0.75}
\definecolor{gray76}{rgb}{0.76,0.76,0.76}
\definecolor{gray77}{rgb}{0.77,0.77,0.77}
\definecolor{gray78}{rgb}{0.78,0.78,0.78}
\definecolor{gray79}{rgb}{0.79,0.79,0.79}
\definecolor{gray7}{rgb}{0.07,0.07,0.07}
\definecolor{gray80}{rgb}{0.80,0.80,0.80}
\definecolor{gray81}{rgb}{0.81,0.81,0.81}
\definecolor{gray82}{rgb}{0.82,0.82,0.82}
\definecolor{gray83}{rgb}{0.83,0.83,0.83}
\definecolor{gray84}{rgb}{0.84,0.84,0.84}
\definecolor{gray85}{rgb}{0.85,0.85,0.85}
\definecolor{gray86}{rgb}{0.86,0.86,0.86}
\definecolor{gray87}{rgb}{0.87,0.87,0.87}
\definecolor{gray88}{rgb}{0.88,0.88,0.88}
\definecolor{gray89}{rgb}{0.89,0.89,0.89}
\definecolor{gray8}{rgb}{0.08,0.08,0.08}
\definecolor{gray90}{rgb}{0.90,0.90,0.90}
\definecolor{gray91}{rgb}{0.91,0.91,0.91}
\definecolor{gray92}{rgb}{0.92,0.92,0.92}
\definecolor{gray93}{rgb}{0.93,0.93,0.93}
\definecolor{gray94}{rgb}{0.94,0.94,0.94}
\definecolor{gray95}{rgb}{0.95,0.95,0.95}
\definecolor{gray96}{rgb}{0.96,0.96,0.96}
\definecolor{gray97}{rgb}{0.97,0.97,0.97}
\definecolor{gray98}{rgb}{0.98,0.98,0.98}
\definecolor{gray99}{rgb}{0.99,0.99,0.99}
\definecolor{gray9}{rgb}{0.09,0.09,0.09}
\definecolor{gray}{rgb}{0.75,0.75,0.75}
\definecolor{green1}{rgb}{0.00,1.00,0.00}
\definecolor{green2}{rgb}{0.00,0.93,0.00}
\definecolor{green3}{rgb}{0.00,0.80,0.00}
\definecolor{green4}{rgb}{0.00,0.55,0.00}
\definecolor{greenyellow}{rgb}{0.68,1.00,0.18}
\definecolor{green}{rgb}{0.00,1.00,0.00}
\definecolor{grey0}{rgb}{0.00,0.00,0.00}
\definecolor{grey100}{rgb}{1.00,1.00,1.00}
\definecolor{grey10}{rgb}{0.10,0.10,0.10}
\definecolor{grey11}{rgb}{0.11,0.11,0.11}
\definecolor{grey12}{rgb}{0.12,0.12,0.12}
\definecolor{grey13}{rgb}{0.13,0.13,0.13}
\definecolor{grey14}{rgb}{0.14,0.14,0.14}
\definecolor{grey15}{rgb}{0.15,0.15,0.15}
\definecolor{grey16}{rgb}{0.16,0.16,0.16}
\definecolor{grey17}{rgb}{0.17,0.17,0.17}
\definecolor{grey18}{rgb}{0.18,0.18,0.18}
\definecolor{grey19}{rgb}{0.19,0.19,0.19}
\definecolor{grey1}{rgb}{0.01,0.01,0.01}
\definecolor{grey20}{rgb}{0.20,0.20,0.20}
\definecolor{grey21}{rgb}{0.21,0.21,0.21}
\definecolor{grey22}{rgb}{0.22,0.22,0.22}
\definecolor{grey23}{rgb}{0.23,0.23,0.23}
\definecolor{grey24}{rgb}{0.24,0.24,0.24}
\definecolor{grey25}{rgb}{0.25,0.25,0.25}
\definecolor{grey26}{rgb}{0.26,0.26,0.26}
\definecolor{grey27}{rgb}{0.27,0.27,0.27}
\definecolor{grey28}{rgb}{0.28,0.28,0.28}
\definecolor{grey29}{rgb}{0.29,0.29,0.29}
\definecolor{grey2}{rgb}{0.02,0.02,0.02}
\definecolor{grey30}{rgb}{0.30,0.30,0.30}
\definecolor{grey31}{rgb}{0.31,0.31,0.31}
\definecolor{grey32}{rgb}{0.32,0.32,0.32}
\definecolor{grey33}{rgb}{0.33,0.33,0.33}
\definecolor{grey34}{rgb}{0.34,0.34,0.34}
\definecolor{grey35}{rgb}{0.35,0.35,0.35}
\definecolor{grey36}{rgb}{0.36,0.36,0.36}
\definecolor{grey37}{rgb}{0.37,0.37,0.37}
\definecolor{grey38}{rgb}{0.38,0.38,0.38}
\definecolor{grey39}{rgb}{0.39,0.39,0.39}
\definecolor{grey3}{rgb}{0.03,0.03,0.03}
\definecolor{grey40}{rgb}{0.40,0.40,0.40}
\definecolor{grey41}{rgb}{0.41,0.41,0.41}
\definecolor{grey42}{rgb}{0.42,0.42,0.42}
\definecolor{grey43}{rgb}{0.43,0.43,0.43}
\definecolor{grey44}{rgb}{0.44,0.44,0.44}
\definecolor{grey45}{rgb}{0.45,0.45,0.45}
\definecolor{grey46}{rgb}{0.46,0.46,0.46}
\definecolor{grey47}{rgb}{0.47,0.47,0.47}
\definecolor{grey48}{rgb}{0.48,0.48,0.48}
\definecolor{grey49}{rgb}{0.49,0.49,0.49}
\definecolor{grey4}{rgb}{0.04,0.04,0.04}
\definecolor{grey50}{rgb}{0.50,0.50,0.50}
\definecolor{grey51}{rgb}{0.51,0.51,0.51}
\definecolor{grey52}{rgb}{0.52,0.52,0.52}
\definecolor{grey53}{rgb}{0.53,0.53,0.53}
\definecolor{grey54}{rgb}{0.54,0.54,0.54}
\definecolor{grey55}{rgb}{0.55,0.55,0.55}
\definecolor{grey56}{rgb}{0.56,0.56,0.56}
\definecolor{grey57}{rgb}{0.57,0.57,0.57}
\definecolor{grey58}{rgb}{0.58,0.58,0.58}
\definecolor{grey59}{rgb}{0.59,0.59,0.59}
\definecolor{grey5}{rgb}{0.05,0.05,0.05}
\definecolor{grey60}{rgb}{0.60,0.60,0.60}
\definecolor{grey61}{rgb}{0.61,0.61,0.61}
\definecolor{grey62}{rgb}{0.62,0.62,0.62}
\definecolor{grey63}{rgb}{0.63,0.63,0.63}
\definecolor{grey64}{rgb}{0.64,0.64,0.64}
\definecolor{grey65}{rgb}{0.65,0.65,0.65}
\definecolor{grey66}{rgb}{0.66,0.66,0.66}
\definecolor{grey67}{rgb}{0.67,0.67,0.67}
\definecolor{grey68}{rgb}{0.68,0.68,0.68}
\definecolor{grey69}{rgb}{0.69,0.69,0.69}
\definecolor{grey6}{rgb}{0.06,0.06,0.06}
\definecolor{grey70}{rgb}{0.70,0.70,0.70}
\definecolor{grey71}{rgb}{0.71,0.71,0.71}
\definecolor{grey72}{rgb}{0.72,0.72,0.72}
\definecolor{grey73}{rgb}{0.73,0.73,0.73}
\definecolor{grey74}{rgb}{0.74,0.74,0.74}
\definecolor{grey75}{rgb}{0.75,0.75,0.75}
\definecolor{grey76}{rgb}{0.76,0.76,0.76}
\definecolor{grey77}{rgb}{0.77,0.77,0.77}
\definecolor{grey78}{rgb}{0.78,0.78,0.78}
\definecolor{grey79}{rgb}{0.79,0.79,0.79}
\definecolor{grey7}{rgb}{0.07,0.07,0.07}
\definecolor{grey80}{rgb}{0.80,0.80,0.80}
\definecolor{grey81}{rgb}{0.81,0.81,0.81}
\definecolor{grey82}{rgb}{0.82,0.82,0.82}
\definecolor{grey83}{rgb}{0.83,0.83,0.83}
\definecolor{grey84}{rgb}{0.84,0.84,0.84}
\definecolor{grey85}{rgb}{0.85,0.85,0.85}
\definecolor{grey86}{rgb}{0.86,0.86,0.86}
\definecolor{grey87}{rgb}{0.87,0.87,0.87}
\definecolor{grey88}{rgb}{0.88,0.88,0.88}
\definecolor{grey89}{rgb}{0.89,0.89,0.89}
\definecolor{grey8}{rgb}{0.08,0.08,0.08}
\definecolor{grey90}{rgb}{0.90,0.90,0.90}
\definecolor{grey91}{rgb}{0.91,0.91,0.91}
\definecolor{grey92}{rgb}{0.92,0.92,0.92}
\definecolor{grey93}{rgb}{0.93,0.93,0.93}
\definecolor{grey94}{rgb}{0.94,0.94,0.94}
\definecolor{grey95}{rgb}{0.95,0.95,0.95}
\definecolor{grey96}{rgb}{0.96,0.96,0.96}
\definecolor{grey97}{rgb}{0.97,0.97,0.97}
\definecolor{grey98}{rgb}{0.98,0.98,0.98}
\definecolor{grey99}{rgb}{0.99,0.99,0.99}
\definecolor{grey9}{rgb}{0.09,0.09,0.09}
\definecolor{grey}{rgb}{0.75,0.75,0.75}
\definecolor{honeydew1}{rgb}{0.94,1.00,0.94}
\definecolor{honeydew2}{rgb}{0.88,0.93,0.88}
\definecolor{honeydew3}{rgb}{0.76,0.80,0.76}
\definecolor{honeydew4}{rgb}{0.51,0.55,0.51}
\definecolor{honeydew}{rgb}{0.94,1.00,0.94}
\definecolor{hotpink}{rgb}{1.00,0.41,0.71}
\definecolor{indianred}{rgb}{0.80,0.36,0.36}
\definecolor{ivory1}{rgb}{1.00,1.00,0.94}
\definecolor{ivory2}{rgb}{0.93,0.93,0.88}
\definecolor{ivory3}{rgb}{0.80,0.80,0.76}
\definecolor{ivory4}{rgb}{0.55,0.55,0.51}
\definecolor{ivory}{rgb}{1.00,1.00,0.94}
\definecolor{khaki1}{rgb}{1.00,0.96,0.56}
\definecolor{khaki2}{rgb}{0.93,0.90,0.52}
\definecolor{khaki3}{rgb}{0.80,0.78,0.45}
\definecolor{khaki4}{rgb}{0.55,0.53,0.31}
\definecolor{khaki}{rgb}{0.94,0.90,0.55}
\definecolor{lavenderblush}{rgb}{1.00,0.94,0.96}
\definecolor{lavender}{rgb}{0.90,0.90,0.98}
\definecolor{lawngreen}{rgb}{0.49,0.99,0.00}
\definecolor{lemonchiffon}{rgb}{1.00,0.98,0.80}
\definecolor{lightblue}{rgb}{0.68,0.85,0.90}
\definecolor{lightcoral}{rgb}{0.94,0.50,0.50}
\definecolor{lightcyan}{rgb}{0.88,1.00,1.00}
\definecolor{lightgoldenrod}{rgb}{0.93,0.87,0.51}
\definecolor{lightgoldenrod}{rgb}{0.98,0.98,0.82}
\definecolor{lightgray}{rgb}{0.83,0.83,0.83}
\definecolor{lightgreen}{rgb}{0.56,0.93,0.56}
\definecolor{lightgrey}{rgb}{0.83,0.83,0.83}
\definecolor{lightpink}{rgb}{1.00,0.71,0.76}
\definecolor{lightsalmon}{rgb}{1.00,0.63,0.48}
\definecolor{lightsea}{rgb}{0.13,0.70,0.67}
\definecolor{lightsky}{rgb}{0.53,0.81,0.98}
\definecolor{lightslate}{rgb}{0.47,0.53,0.60}
\definecolor{lightslate}{rgb}{0.47,0.53,0.60}
\definecolor{lightslate}{rgb}{0.52,0.44,1.00}
\definecolor{lightsteel}{rgb}{0.69,0.77,0.87}
\definecolor{lightyellow}{rgb}{1.00,1.00,0.88}
\definecolor{limegreen}{rgb}{0.20,0.80,0.20}
\definecolor{linen}{rgb}{0.98,0.94,0.90}
\definecolor{magenta1}{rgb}{1.00,0.00,1.00}
\definecolor{magenta2}{rgb}{0.93,0.00,0.93}
\definecolor{magenta3}{rgb}{0.80,0.00,0.80}
\definecolor{magenta4}{rgb}{0.55,0.00,0.55}
\definecolor{magenta}{rgb}{1.00,0.00,1.00}
\definecolor{maroon1}{rgb}{1.00,0.20,0.70}
\definecolor{maroon2}{rgb}{0.93,0.19,0.65}
\definecolor{maroon3}{rgb}{0.80,0.16,0.56}
\definecolor{maroon4}{rgb}{0.55,0.11,0.38}
\definecolor{maroon}{rgb}{0.69,0.19,0.38}
\definecolor{mediumaquamarine}{rgb}{0.40,0.80,0.67}
\definecolor{mediumblue}{rgb}{0.00,0.00,0.80}
\definecolor{mediumorchid}{rgb}{0.73,0.33,0.83}
\definecolor{mediumpurple}{rgb}{0.58,0.44,0.86}
\definecolor{mediumsea}{rgb}{0.24,0.70,0.44}
\definecolor{mediumslate}{rgb}{0.48,0.41,0.93}
\definecolor{mediumspring}{rgb}{0.00,0.98,0.60}
\definecolor{mediumturquoise}{rgb}{0.28,0.82,0.80}
\definecolor{mediumviolet}{rgb}{0.78,0.08,0.52}
\definecolor{midnightblue}{rgb}{0.10,0.10,0.44}
\definecolor{mintcream}{rgb}{0.96,1.00,0.98}
\definecolor{mistyrose}{rgb}{1.00,0.89,0.88}
\definecolor{moccasin}{rgb}{1.00,0.89,0.71}
\definecolor{navajowhite}{rgb}{1.00,0.87,0.68}
\definecolor{navyblue}{rgb}{0.00,0.00,0.50}
\definecolor{navy}{rgb}{0.00,0.00,0.50}
\definecolor{oldlace}{rgb}{0.99,0.96,0.90}
\definecolor{olivedrab}{rgb}{0.42,0.56,0.14}
\definecolor{orange1}{rgb}{1.00,0.65,0.00}
\definecolor{orange2}{rgb}{0.93,0.60,0.00}
\definecolor{orange3}{rgb}{0.80,0.52,0.00}
\definecolor{orange4}{rgb}{0.55,0.35,0.00}
\definecolor{orangered}{rgb}{1.00,0.27,0.00}
\definecolor{orange}{rgb}{1.00,0.65,0.00}
\definecolor{orchid1}{rgb}{1.00,0.51,0.98}
\definecolor{orchid2}{rgb}{0.93,0.48,0.91}
\definecolor{orchid3}{rgb}{0.80,0.41,0.79}
\definecolor{orchid4}{rgb}{0.55,0.28,0.54}
\definecolor{orchid}{rgb}{0.85,0.44,0.84}
\definecolor{palegoldenrod}{rgb}{0.93,0.91,0.67}
\definecolor{palegreen}{rgb}{0.60,0.98,0.60}
\definecolor{paleturquoise}{rgb}{0.69,0.93,0.93}
\definecolor{paleviolet}{rgb}{0.86,0.44,0.58}
\definecolor{papayawhip}{rgb}{1.00,0.94,0.84}
\definecolor{peachpuff}{rgb}{1.00,0.85,0.73}
\definecolor{peru}{rgb}{0.80,0.52,0.25}
\definecolor{pink1}{rgb}{1.00,0.71,0.77}
\definecolor{pink2}{rgb}{0.93,0.66,0.72}
\definecolor{pink3}{rgb}{0.80,0.57,0.62}
\definecolor{pink4}{rgb}{0.55,0.39,0.42}
\definecolor{pink}{rgb}{1.00,0.75,0.80}
\definecolor{plum1}{rgb}{1.00,0.73,1.00}
\definecolor{plum2}{rgb}{0.93,0.68,0.93}
\definecolor{plum3}{rgb}{0.80,0.59,0.80}
\definecolor{plum4}{rgb}{0.55,0.40,0.55}
\definecolor{plum}{rgb}{0.87,0.63,0.87}
\definecolor{powderblue}{rgb}{0.69,0.88,0.90}
\definecolor{purple1}{rgb}{0.61,0.19,1.00}
\definecolor{purple2}{rgb}{0.57,0.17,0.93}
\definecolor{purple3}{rgb}{0.49,0.15,0.80}
\definecolor{purple4}{rgb}{0.33,0.10,0.55}
\definecolor{purple}{rgb}{0.63,0.13,0.94}
\definecolor{red1}{rgb}{1.00,0.00,0.00}
\definecolor{red2}{rgb}{0.93,0.00,0.00}
\definecolor{red3}{rgb}{0.80,0.00,0.00}
\definecolor{red4}{rgb}{0.55,0.00,0.00}
\definecolor{red}{rgb}{1.00,0.00,0.00}
\definecolor{rosybrown}{rgb}{0.74,0.56,0.56}
\definecolor{royalblue}{rgb}{0.25,0.41,0.88}
\definecolor{saddlebrown}{rgb}{0.55,0.27,0.07}
\definecolor{salmon1}{rgb}{1.00,0.55,0.41}
\definecolor{salmon2}{rgb}{0.93,0.51,0.38}
\definecolor{salmon3}{rgb}{0.80,0.44,0.33}
\definecolor{salmon4}{rgb}{0.55,0.30,0.22}
\definecolor{salmon}{rgb}{0.98,0.50,0.45}
\definecolor{sandybrown}{rgb}{0.96,0.64,0.38}
\definecolor{seagreen}{rgb}{0.18,0.55,0.34}
\definecolor{seashell1}{rgb}{1.00,0.96,0.93}
\definecolor{seashell2}{rgb}{0.93,0.90,0.87}
\definecolor{seashell3}{rgb}{0.80,0.77,0.75}
\definecolor{seashell4}{rgb}{0.55,0.53,0.51}
\definecolor{seashell}{rgb}{1.00,0.96,0.93}
\definecolor{sienna1}{rgb}{1.00,0.51,0.28}
\definecolor{sienna2}{rgb}{0.93,0.47,0.26}
\definecolor{sienna3}{rgb}{0.80,0.41,0.22}
\definecolor{sienna4}{rgb}{0.55,0.28,0.15}
\definecolor{sienna}{rgb}{0.63,0.32,0.18}
\definecolor{skyblue}{rgb}{0.53,0.81,0.92}
\definecolor{slateblue}{rgb}{0.42,0.35,0.80}
\definecolor{slategray}{rgb}{0.44,0.50,0.56}
\definecolor{slategrey}{rgb}{0.44,0.50,0.56}
\definecolor{snow1}{rgb}{1.00,0.98,0.98}
\definecolor{snow2}{rgb}{0.93,0.91,0.91}
\definecolor{snow3}{rgb}{0.80,0.79,0.79}
\definecolor{snow4}{rgb}{0.55,0.54,0.54}
\definecolor{snow}{rgb}{1.00,0.98,0.98}
\definecolor{springgreen}{rgb}{0.00,1.00,0.50}
\definecolor{steelblue}{rgb}{0.27,0.51,0.71}
\definecolor{tan1}{rgb}{1.00,0.65,0.31}
\definecolor{tan2}{rgb}{0.93,0.60,0.29}
\definecolor{tan3}{rgb}{0.80,0.52,0.25}
\definecolor{tan4}{rgb}{0.55,0.35,0.17}
\definecolor{tan}{rgb}{0.82,0.71,0.55}
\definecolor{thistle1}{rgb}{1.00,0.88,1.00}
\definecolor{thistle2}{rgb}{0.93,0.82,0.93}
\definecolor{thistle3}{rgb}{0.80,0.71,0.80}
\definecolor{thistle4}{rgb}{0.55,0.48,0.55}
\definecolor{thistle}{rgb}{0.85,0.75,0.85}
\definecolor{tomato1}{rgb}{1.00,0.39,0.28}
\definecolor{tomato2}{rgb}{0.93,0.36,0.26}
\definecolor{tomato3}{rgb}{0.80,0.31,0.22}
\definecolor{tomato4}{rgb}{0.55,0.21,0.15}
\definecolor{tomato}{rgb}{1.00,0.39,0.28}
\definecolor{turquoise1}{rgb}{0.00,0.96,1.00}
\definecolor{turquoise2}{rgb}{0.00,0.90,0.93}
\definecolor{turquoise3}{rgb}{0.00,0.77,0.80}
\definecolor{turquoise4}{rgb}{0.00,0.53,0.55}
\definecolor{turquoise}{rgb}{0.25,0.88,0.82}
\definecolor{violetred}{rgb}{0.82,0.13,0.56}
\definecolor{violet}{rgb}{0.93,0.51,0.93}
\definecolor{wheat1}{rgb}{1.00,0.91,0.73}
\definecolor{wheat2}{rgb}{0.93,0.85,0.68}
\definecolor{wheat3}{rgb}{0.80,0.73,0.59}
\definecolor{wheat4}{rgb}{0.55,0.49,0.40}
\definecolor{wheat}{rgb}{0.96,0.87,0.70}
\definecolor{whitesmoke}{rgb}{0.96,0.96,0.96}
\definecolor{white}{rgb}{1.00,1.00,1.00}
\definecolor{yellow1}{rgb}{1.00,1.00,0.00}
\definecolor{yellow2}{rgb}{0.93,0.93,0.00}
\definecolor{yellow3}{rgb}{0.80,0.80,0.00}
\definecolor{yellow4}{rgb}{0.55,0.55,0.00}
\definecolor{yellowgreen}{rgb}{0.60,0.80,0.20}
\definecolor{yellow}{rgb}{1.00,1.00,0.00}

\usepackage{natbib}
\usepackage{graphicx}

\begin{document}


\title{Identifying the progenitor set of present-day early-type galaxies: a view from the standard model}


\author{Sugata Kaviraj \inst{1,2} \thanks{skaviraj@astro.ox.ac.uk; s.kaviraj@imperial.ac.uk} \and Julien Devriendt \inst{2,3} \and Ignacio Ferreras \inst{1} \and Sukyoung Yi \inst{4} \and Joseph Silk \inst{2}}

\institute{Mullard Space Science Laboratory, Holmbury St. Mary,
Dorking, Surrey RH5 6NT UK \and Department of Physics, University
of Oxford, Keble Road, Oxford OX1 3RH, UK \and Observatoire
Astronomique de Lyon, 9 Avenue Charles Andr\'e, 69561 Saint-Genis
Laval cedex, France \and Center for Space Astrophysics, Yonsei
University, 134 Shinchon, Seoul 120-749, Korea}


\date{Accepted for publication in A\&A}


\abstract{ {\color{black}We present a comprehensive theoretical
study, using a semi-analytical model within the standard LCDM
framework, of the photometric properties of the progenitors of
present-day early-type galaxies in the redshift range $0<z<1$. We
explore progenitors of \emph{all morphologies} and study their
characteristics as a function of the luminosity and local
environment of the early-type remnant at $z=0$. In agreement with
previous studies, we find that, while larger early-types are
generally assembled later, their luminosity-weighted stellar ages
are typically older.} In dense cluster-like environments, $\sim70$
percent of early-type systems are `in place' by $z=1$
{\color{black}and evolve without interactions thereafter}, while
in the field the corresponding value is $\sim30$ percent.
Averaging across all environments at $z\sim1$, less than 50
percent of the stellar mass which ends up in early-types today is
actually in early-type progenitors at this redshift,
{\color{black}in agreement with recent observational work.} The
corresponding value is $\sim65$ percent in clusters, due to faster
morphological evolution in such dense environments.
{\color{black}We develop probabilistic prescriptions which provide
a means of including spiral (i.e. non early-type) progenitors at
intermediate and high redshifts, based on their luminosity and
optical colours}. For example, we find that, at intermediate
redshifts ($z\sim0.5$), large ($M_V<-21.5$), red ($B-V>0.7$)
spirals have $\sim75-95$ percent chance of being an early-type
progenitor, while the corresponding probability for large blue
spirals ($M_B<-21.5$, $B-V<0.7$) is $\sim50-75$ percent.
{\color{black}The prescriptions developed here can be used to
address, from the perspective of the standard model, the issue of
`progenitor bias', whereby the exclusion of late-type progenitors
in observational studies can lead to inaccurate conclusions
regarding the evolution of the early-type population over cosmic
time.} Finally, we explore the correspondence between the true
`progenitor set' of the present-day early-type population -
defined as the set of all galaxies that are progenitors of
present-day early-types regardless of their morphologies - and the
frequently used `red-sequence', defined as the set of galaxies
within the part of the colour-magnitude space which is dominated
by early-type objects. {\color{black}We find that, while more
massive members ($M_V\leq-21$) of the `red sequence' trace the
progenitor set reasonably well, the relationship breaks down at
fainter luminosities ($M_V\geq-21$). Thus, while the results of
recent observational studies which exploit the red sequence are
valid (since they are largely restricted to massive galaxies),
more care should be taken when deeper observations (which will
probe fainter luminosities) become available in the future.}}



\authorrunning{Sugata Kaviraj et al.}
\titlerunning{Early-type progenitors in the standard model}
\maketitle


\section{Introduction}
As `end points' of galaxy merger sequences, early-type galaxies
carry important signatures of mass assembly and star formation in
the Universe. Deducing their star formation histories (SFHs)
therefore contains the key to understanding not only the evolution
of these galaxies but the evolutionary patterns of galaxies as a
whole. Our view of early-type galaxy formation has developed over
the years, away from the classical `monolithic collapse'
hypothesis \citep[e.g.][]{Larson1975,Chiosi2002} and towards the
hierarchical assembly of these objects through mergers and
accretion of smaller galaxies over time, in the framework of the
currently popular LCDM paradigm of galaxy formation
\citep[e.g.][and references
therein]{KWG93,KCW96,Baugh1996,KC98,SP99,Cole2000,Hatton2003,deLucia2006,Bower2006}.

{\color{black}A significant body of observational work in the past
has traced the assembly histories of early-type galaxies by
studying only early-type populations at high redshift
\citep[e.g][]{Gladders1998,Stanford1998,Barrientos2003,Ferreras2005}.}
However, a fundamental feature of early-type formation in the
standard LCDM model is that stellar mass that eventually ends up
in present-day early-type galaxies is not entirely contained in
early-type systems at high redshift. Although, early-types at any
redshift are almost guaranteed progenitors of their counterparts
at present-day, looking only at early-types at high redshift
introduces a `progenitor bias', which becomes increasingly more
severe at larger look-back times, as the fraction of early-type
galaxies becomes progressively smaller and late-type systems begin
to dominate the progenitor population
{\color{black}\citep[e.g.][]{Franx1996,VD2000,VD2001,Kaviraj2005a}}.

In the current era of large scale surveys, e.g. SDSS
(Adelman-McCarthy et al. 2006), COMBO-17 (Wolf et al. 2004), MUSYC
(Gawiser et al. 2006), GEMS (Rix et al. 2004), unprecedented
amounts of data spanning a large range in redshift (typically
$0<z<1$) and environment are becoming available, allowing us to
study statistically significant numbers of galaxies at various
stages of evolution. {\color{black}A quantitative study of
early-type progenitors within the standard model is therefore
desirable to (a) understand the difference in the way early-types
are assembled as a function of their luminosity and environment
and (b) to gauge the role of non-early-type progenitors in
early-type evolution, especially for studies that focus at high
redshift.}

{\color{black}A central theme of this work is to use a model which
accurately reproduces the photometric properties of early-type
galaxies and their observed evolution over a large range in
redshift to study, in detail, the properties of the progenitors of
early-type galaxies at present day. The process of mapping the
progenitor population and its evolution with redshift also
provides a realistic picture of progenitor bias within the
framework of the standard model.}

{\color{black}While \citet{VD2001} have developed and studied the
issue of progenitor bias, their study employed phenomenological
SFHs, with the simple (and perhaps unrealistic) assumption that
morphological transformations occur abruptly $\sim$ 1.5 Gyrs after
the cessation of star formation in a particular galaxy. Their work
was an extension of previous ideas
\citep[e.g][]{Bower98,Shioya1998} that took into account the
potential for elliptical galaxies to have complex star formation
histories, but did not explore the effect of morphological
transformations on the evolution of the early-type population at
high redshift.}

{\color{black}While some recent observational studies
\citep[e.g.][]{Holden2005,vandeVen2003,VD2007} have used the
results of \citet{VD2001} to take the effects of progenitor bias
into account, other studies (e.g. Blakeslee et al. 2003) have used
the entire galaxy population, without reference to morphology, in
an attempt to reduce the bias in their conclusions that would
occur if only elliptical galaxies were used in the analysis.}

Our study refines and extends the results of \citet{VD2001} by
studying early-type progenitors within a realistic and
well-calibrated semi-analytical framework, in which mass assembly
and morphological transformations can be followed more accurately
in the context of the LCDM paradigm. {\color{black}While the
\citet{VD2001} study does attempt to alleviate the effects of
progenitor bias, its simplicity, especially in the prescription
used for morphological transformations, makes a more realistic
treatment of this issue, in the framework of the standard model,
very desirable. Using the entire galaxy population without
reference to morphology is also not ideal because that implies
that all galaxies at high redshift are potential progenitors of
present-day early-type galaxies. While this approximation may be
reasonably robust for the most massive early-type galaxies in the
densest regions of the Universe, our study shows that this is
generally not applicable.}

{\color{black} In this paper we study the evolution of the
`progenitor set' of early-type galaxies, defined as the set of all
galaxies at a given redshift that are, regardless of morphology,
the progenitors of an early-type at $z=0$}. {\color{black}Note
that `early-types' are defined as galaxies that are elliptical or
lenticular. See Section 2 for a description of how galaxy
morphologies are defined in the model.} We explore the evolution
of the progenitor set with redshift, as a function of the
luminosity and environment of the early-type remnant which is left
at present-day. We pay particular attention to \emph{spiral
progenitors} in the model, since these might have been excluded
from some studies of early-type evolution in the past, by virtue
of their morphology, even though they form an important part of
the progenitor set. By comparing the properties (optical colours
and luminosities) of spiral progenitors to the general spiral
population, we provide a means of correcting for progenitor bias,
which is consistent with the properties of the standard model, by
including specific parts of the spiral population at high redshift
into the study of early-type evolution.

The plan of this paper is as follows. {\color{black}In Section 2,
we describe the salient features of the model used in this study}.
In Section 3, we quantify the morphologies of the galaxies that
make up the progenitor set. We map the properties of elliptical,
S0 and spiral progenitors as a function of redshift and explore
differences between progenitors as a function of the mass and
environment of the elliptical remnant at present day. In Section 4
we focus exclusively on spiral progenitors and compare their
photometric properties to the general spiral population, to derive
probabilistic prescriptions for including spiral progenitors in
early-type studies. Section 5 traces the contribution of galaxies
in dense regions at high redshift to cluster early-types at
present-day. Finally, in Section 6 we explore the correspondence
between the true progenitor set of present-day early-type galaxies
and the `red-sequence', which is sometimes used as a proxy for the
progenitor set in observational studies (e.g. Bell et al. 2004,
Faber et al. 2007).

Note that throughout this study we provide rest-frame magnitudes
for all model galaxies. Unless otherwise noted, the filters used
are in the standard Johnson system.


{\color{black}
\section{The model}
The semi-analytical model used in this study is GALICS, which
combines large-scale cosmological N-body simulations with
analytical recipes for the evolution of baryons within dark matter
haloes \citep{Hatton2003}. GALICS makes predictions for the
overall statistical properties of galaxy populations, with an
emphasis on their panchromatic spectral energy distributions
across a wide wavelength range ($UV$ to infrared/submillimetre).
In this section we describe the salient features of the model,
including the implementation of the cosmological simulation and
details of the semi-analytics that determine the growth of bulges.
We also highlight the successes of GALICS in terms of reproducing
the photometric properties of early-type galaxies in the
$UV$-optical spectral ranges across a wide redshift range.

{\color{black}Before we begin our analysis we briefly describe the
definition of galaxy morphology in the model used in this study.
Galaxy morphology in the model is determined by the ratio of the
B-band luminosities of the disk and bulge components which
correlates well with Hubble type \citep{Simien1986}. A morphology
index is defined as

\begin{equation}
I = \exp\bigg(\frac{-L_B}{L_D}\bigg)
\end{equation}

such that a pure disk has $I=1$ and a pure bulge has $I=0$.
Following \citet{Baugh1996}, ellipticals have $I<0.219$, S0s have
$0.219<I<0.507$ and spirals have $I>0.507$.}


\subsection{Dark matter simulation}
The dark matter `backbone' used in GALICS has been generated by
simulating a LCDM model in a cube with a comoving size of
100$h^{-1}$Mpc with parameters $\Omega_m=0.33$,
$\Omega_{\Lambda}=0.667$ and $\sigma_8 = 0.88$. The amplitude of
the power spectrum is computed using the present-day abundance of
rich clusters \citep{Eke1996} and initial conditions are extracted
using the GRAFIC code \citep{Bertschinger1995}. The simulation
contains $256^3$ dark matter particles. The mass of each dark
matter particle is $\sim8 \times 10^9M_{\odot}$ and the spatial
resolution achieved is $\sim29.3$ kpc. The minimum mass of
galaxies (due to the minimum number of particles used in the
friend-of-friends group finder and the baryon fraction used) is
$\sim2 \times 10^{10} M_{\odot}$.


\subsection{The growth of bulges in the semi-analytical model}
\subsubsection{Mergers vs disk instabilities}
GALICS includes two processes which lead to the formation and
growth of spheroidal components in galaxies: mergers (both major
and minor) and gravitational disk instabilities
{\color{black}\footnote{A disk is considered to be stable if
$V_c<0.7V_{tot}$, where $V_c$ is the circular velocity of the disk
and $V_{tot}$ is the rotational velocity of the disk-bulge-halo
system \citep[e.g.][]{vandenbosch1998}.}}. While mergers directly
reveal the hierarchical nature of galaxy formation, disk
instabilities can result in the formation of bulges in a
`monolithic' manner (i.e. in objects classified as isolated at our
resolution limit). The relative contribution of each of these two
processes in the build-up of bright, local galaxy bulges thus
yields a natural estimate of how hierarchical galaxy formation
truly is.

For each of our galaxies we measure the stellar and gas masses
transfered from the disk to the bulge during mergers ($M_{merg}$)
and/or disk instabilities ($M_{inst}$). In Figure
\ref{fig:merg_over_inst}, we show the relative contributions of
mergers (both major and minor) and disk instabilities to the mass
build-up of the spheroidal components of local galaxies brighter
than $M(B)=-18.9$. This is shown as the distribution of the ratio
$M_{merg}/M_{inst}$, computed for each model galaxy. A galaxy with
a value of 1 for this ratio has acquired half its bulge mass
during mergers and the other half through disk instabilities. Note
that we only consider galaxies with a bulge component. Pure disk
galaxies (i.e. those that have never been unstable or undergone a
merger) represent only $\sim 1.4$ percent of the bright
population.

Several important conclusions can be drawn from this figure.
First, it is clear that the majority of elliptical galaxies
develop their bulges during mergers rather than disk
instabilities. This result may seem natural, as GALICS is an
implementation of the hierarchical galaxy formation scenario.
However, the main point here is that, for the first time, we do
indeed compare the relative contribution of mergers to other
processes in the formation of spheroids. The fact that model
elliptical galaxies develop their bulges mainly through mergers
rather than through disk instabilities is an \emph{output} of our
model - we do not simply assume that elliptical galaxies only form
during mergers but instead allow the different physical processes
we believe are relevant to the formation process to compete. In
that sense, one can consider the results we obtain as a
demonstration that the formation of ellipticals is indeed
hierarchical.

Second, elliptical and lenticular galaxy formation appears to be
somewhat different, since GALICS predicts that most lenticular
galaxies develop their bulges through disk instabilities. Changing
the modelling of disk instabilities would affect the plots in a
similar way for ellipticals and lenticulars, leaving the
difference in their respective behaviours unchanged. The fact that
the model lenticulars acquire their morphology through disk
instabilities is a result which is common in numerical simulations
of galactic dynamics \citep{Combes2000} and gives us some
confidence that our simple physical modelling broadly captures the
mechanism.

Finally, the lower panels of Figure \ref{fig:merg_over_inst}
indicate that spirals form their bulge components mainly through
disk instabilities. The global picture for morphogenesis that we
obtain from GALICS is satisfactory as it compares well to more
detailed numerical simulations. Moreover, we show that whilst an
`isolated object' approximation can be justified to describe
spiral galaxies (and some lenticulars), it cannot be used to
properly model the formation of the vast majority of ellipticals
in the framework of CDM structure formation.


{\color{black}\subsubsection{A note on the treatment of disk
instabilities} The disk instability model employed is similar to
that of van den Bosch (1998). Rotational equilibrium is enforced
at the disk half mass radius - in other words, the circular
velocity $V_{tot}$ arising from the presence of total mass of gas,
stars and dark matter enclosed in the half mass radius sphere is
calculated. It is then compared to the circular velocity $V_c$ of
the disk alone, and if this latter is smaller than a critical
value (obtained by mass weighting the value of 0.7 for a pure gas
disk and 0.52 for a pure stellar disk), the minimum amount of mass
of gas and stars required to re-establish stability is transferred
to the disk.}


\subsubsection{Post-merger morphology}
In the literature mergers are typically modelled by taking the
ratio of the progenitors, adding the stars of the lighter galaxy
to the disk of the heavier one if the mass ratio is less than
$f_{bulge} \sim 0.3$ or destroy the disk and form a bulge if the
ratio is higher
\citep[e.g.][]{Walker1996,Cole2000,SP99,Kauffmann1999}{\color{black}\footnote{In
keeping with the literature, we define a merger as `major' when
the mass ratio between merging objects ranges between one and a
third, and as `minor' when it is smaller.}}. Since a galactic disk
can be disrupted by an encounter even when the interloper is less
massive than the disk itself, this simple prescription reproduces
this behaviour. However, the sharp cut-off at $f_{bulge} \sim 0.3$
between totally disrupting or not disrupting the morphology of the
galaxy seems somewhat unrealistic.

The model implemented in GALICS is constructed in terms of a
smooth function ($X$), which models the fraction of disk material
remaining in the disk after the merger as a function of the mass
ratio of the two progenitors. The previous studies (described
above) effectively use a step function, in the sense that the
material kept in the disk is either 0 or 1 depending on whether
the mass ratio is greater or lesser than 0.3 respectively. In
GALICS this smooth function is defined as:

\begin{equation}
X(R) = \Bigg[1+\Bigg(\frac{\chi-1}{R-1}\Bigg)^\chi\Bigg]^{-1},
\end{equation}

where $R$ represents the mass ratio of the heavier to the lighter
progenitor and $\chi$ is the critical value of that ratio i.e. the
value that $X=0.5$. Since step functions with $f_{bulge}\sim0.3$
have been found to give good results in the past, the fiducial
value adopted in GALICS is $\chi=1/0.3$. See Figure 3 in
\citet{Hatton2003} for a visualisation of $X(R)$ for this fiducial
value of $\chi$. We refer readers to Section 5 in
\citet{Hatton2003} for details of prescriptions related to merging
employed by GALICS.

\begin{figure}
\begin{center}
\includegraphics[width=3.5in]{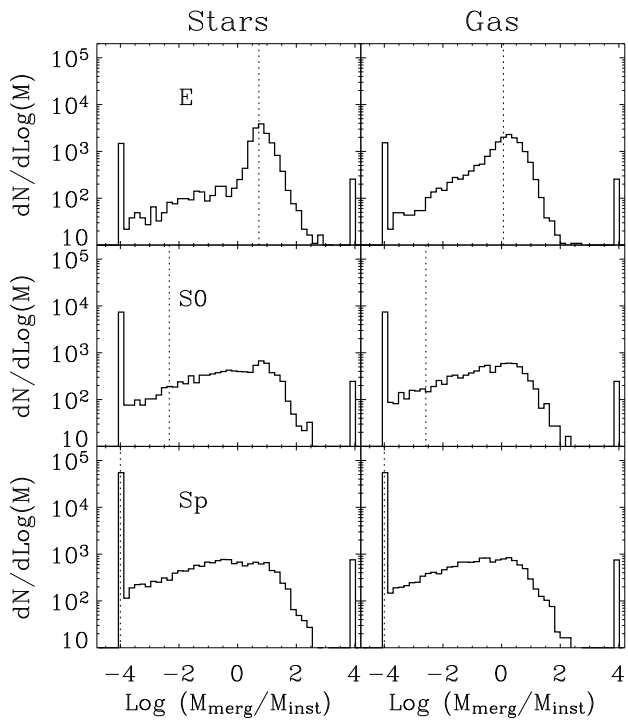}\\
\caption{The ratio between stellar and gas masses transferred from
the disk to the bulge during mergers ($M_{merg}$) and through disk
instabilities ($M_{inst}$). A galaxy with a value of 1 for this
ratio has acquired half its bulge mass during mergers and the
other half through disk instabilities. This plot is restricted to
model galaxies brighter than $M(B)=-18.9$. {\color{black}Note that
the (artificial) peaks at 4 and -4 are for ratios
$M_{merg}/M_{inst} = \infty$ and 0 respectively. The vertical
dashed lines represent median values.}} \label{fig:merg_over_inst}
\end{center}
\end{figure}

{\color{black} Galaxies are modelled with three components: the,
disk, the bulge and the burst. The burst has the same geometry
(and therefore star formation law) as the bulge but its scale
radius is 10\% of that of the bulge. During a merger, a fraction
of $X$ of both gas and stars originally in the disk remain in the
disk while the rest are transferred to the burst. Stars
formed/existing in the burst are rapidly transferred to the bulge.
As the central burst forms stars (typically over short timescales
than the `quiescent' mode due to its smaller scale radius), the
new stellar component gets transferred to the bulge. As it
subsequently runs out of the gas deposited by the merger, the
burst eventually disappears. We refer readers to section 5 of
Hatton et al. (2003) for further details. Note that, for higher
values of $R$ (i.e. mergers with a high mass ratio) almost all
disk material remains in the disk, while equal mass mergers result
in rapid bulge formation, mimicking the prescriptions employed in
the literature.}


{\color{black}
\subsection{A discussion of free parameters in the model}
Inevitably, any study that exploits a semi-analytical model such
as the one presented here, relies on simplified, albeit
well-calibrated recipes to describe galaxy evolution. These
semi-analytic recipes are driven by free parameters which
determine the evolution of the baryonic Universe. While the
parameters that drive the baryonic evolution in the fiducial
GALICS model are discussed in Section 7 of \citet{Hatton2003}, we
briefly revisit them here and discuss their potential impact on
the analysis that follows in the paper. In Table 1 we list the
free parameters in the model and indicate how their values are
constrained in this study. The first section of the table
indicates parameters that have the greatest impact on the
evolution of the galaxy population and the bottom section
indicates parameters that play a minor or negligible role.

While the model is driven by several free parameters, those that
have the greatest impact on the spectro-photometric properties of
the galaxy population are (a) the star formation efficiency (b)
the efficiency of mass loading for supernova feedback and (c) the
merging law that determines morphological transformations during
galaxy interactions. We explore each of these in turn and briefly
discuss the sensitivity of the model to these parameters.

\begin{description}
    \item[\textbf{Star formation efficiency:} ]
The star formation efficiency (SFE) determines the rate at which
cold gas is converted into stars. For example, decreasing the SFE
results in less cold gas being converted into stars, moves the
peak of the star formation rate closer to present day and affects
both the shape and normalisation of the z=0 galaxy luminosity
function. In our fiducial model, the inverse of the SFE ($\beta$)
is set equal to 50 (i.e. the SFE is $\sim$2\%), following
\citet[][see their Eqn. 7]{Kennicutt1998}, who combined H$\alpha$,
HI, CO and far-infrared measurements in spiral and
infrared-selected starburst galaxies to provide a parametrisation
of the global star formation rate in the local Universe. Figure 6
in their study indicates that this parametrisation is accurate
over a wide gas density range, from gas-poor spiral disks to the
cores of the most luminous starburst galaxies, implying that the
value for this parameter that is used in the model is reasonably
well-constrained. Furthermore, the present-day B and K-band galaxy
luminosity functions predicted by GALICS are very consistent with
those observed by the 2dF survey (Cross et al. 2001), indicating
that the effect of the adopted value ($\beta=50$) is consistent
with the observed distribution of galaxy luminosities in the real
Universe.\\

    \item[\textbf{Mass loading for supernova feedback:} ]
For a given stellar mass formed, a certain fraction is contained
in massive, fast-evolving stars that become supernovae (SN),
injecting kinetic energy into the interstellar medium. The
efficiency of this SN driven `wind' depends on both the porosity
of the ISM \citep[see][]{Silk2001} and the `mass loading factor',
which describes the entrainment of interstellar gas by the wind
\citep[see e.g.][]{Silk2001,Silk2003}. For example, increasing the
mass loading efficiency will produce more feedback, heat more cold
gas and reduce the amount of fuel available for star formation.
This will, in turn, affect the galaxy luminosity function at
present-day. Following \citet{Martin2002}, we assume that the mass
loading factor is $\sim$10, which results in the outflow rate of a
starburst to be of the order of the star formation rate (see
Section 4.2 in Hatton et al. (2003) for more details). Thus the
inverse of the mass loading factor ($\epsilon$) is set to 0.1. As
mentioned before, the reproduction of the galaxy luminosity
functions in the B and K bands - which relies in part on
$\epsilon$ - are well reproduced by the fiducial GALICS model.\\

    \item[\textbf{Merger law: }]
A third parameter that has a significant impact on galaxy
evolution is the merger law that determines post-merger
morphology. This law determines the morphological mix of the
Universe at $z=0$ and therefore has a direct impact on the
early-type progenitor set in the model. The prescription used to
determine post-merger morphology has already been described in
detail above in Section 2.2.3. \citet{Hatton2003} indicates that
the predicted morphological mix in GALICS is (within counting
errors) consistent with that seen in the real Universe. Note,
however, that the comparisons presented in Section 8.6 of
\citet{Hatton2003} use reasonably small surveys (e.g. the
Stromlo-APM redshift survey; Loveday et al. 1996) compared to the
scale of modern surveys such as the SDSS. Future papers will
present more robust comparisons between the morphological mix
predicted by GALICS and that observed in the SDSS e.g. through
comparison of the GALICS predictions to visually-inspected
morphologies of the entire SDSS DR6 measured by the `Galaxy Zoo'
project \citep{Lintott2008}.\\

    \item[\textbf{Minor parameters: }]
We complete our description of the free parameters by briefly
discussing minor parameters that do not produce a measurable
impact on the analysis. The normalization for the
satellite-satellite merging law ($\psi$) is constrained using
\citet{Makino1997}. Satellite-satellite merging is much rarer than
dynamical friction merging (i.e. merging with centrals in a halo)
and thus this parameter plays a negligible role in the analysis.

The recycling efficiency ($\varsigma$) describes the fraction of
gas originally expelled from the halo by feedback which is
re-accreted over time as the DM halo grows in mass. We note that
\citet{SP99} set this parameter to 0 (i.e. any gas expelled is
lost from the system altogether) while \citet{Cole2000} set it to
1. Hatton et al. (2003) use a fiducial value of 30\%, to represent
an intermediate behaviour compared to the two extremes described
above, which is plausibly closer to the truth. We note, however,
that the effect of this parameter is very weak because at the mass
resolution of the model, very little gas is expelled from DM
halos. The re-accreted gas is always significantly less than the
amount of `pristine' gas accreted or the amount of gas that is
ejected out of the galaxy and into the halo in the first place.

In GALICS star formation driven by disk instabilities and mergers
takes place in a central `burst' region, which is modelled as
having the same morphology as the bulge but with a fraction of its
radius (see Section 4 in Hatton et al. 2003 for a more details).
The `burst to bulge' radius ($\kappa$) - set to a fiducial value
of 0.1 - dictates the size of the burst region in model galaxies
(i.e. an order of magnitude smaller than the radius of the bulge).
Reducing $\kappa$ would make the burst regions smaller, shortening
the star formation timescales (and vice-versa). However, typical
star formation timescales (a few tens of Myrs at most) in the
burst regions are negligible compared to the timescales over which
galaxy properties are being `viewed' in the model (several Gyrs).
Hence, changing this parameter does not affect the observed
colours of the galaxy population predicted by the model and thus
leaves our conclusions unchanged.

The dust recipe used in the model is calibrated using the Milky
Way, the Large Magellanic Cloud and Small Magellanic Cloud and a
few local spirals for which the extinction curve, gas content and
metallicity have been measured \citep[see][for
details]{Guiderdoni1987}. While a key assumption is that the dust
properties of galaxies are invariant with redshift, tests of this
assumption requires data at high redshift which are not yet
available, leaving us very little room to further calibrate our
dust recipe. Finally, the fiducial GALICS model uses the
\cite{Kennicutt1983} Initial Mass Function (IMF). We note that the
dispersion in the predicted properties of the early-type
population in the local Universe does not vary significantly with
other similar IMFs such as \citet{Salpeter1955} - see Table 1 in
Kaviraj et al. (2005a).
\end{description}

Notwithstanding the large set of free parameters inherent to any
semi-analytical analysis, we note that the fiducial values of the
parameters in the model are calibrated either through
observational data (e.g. the SF efficiency) or through numerical
simulations (e.g. the merger law). Most importantly, the full
suite of calibrated parameters that drives the model predictions
reproduces a fundamental set of spectro-photometric properties of
the observed galaxy population in the Universe. GALICS produces
good agreement to the galaxy luminosity functions observed by the
2dF survey in the B and K bands \citep{Cross2001}. The $(B-V)$
colours of spiral galaxies closely follow the observed data of
\citet{Buta1994}, both in terms of average values and scatter.
Satisfactory fits to the Faber-Jackson relation for early-types
(and the Tully-Fisher relation for disks) are obtained and
predictions show good agreement with the early-type Fundamental
Plane, with the model predicting the observed morphological mix of
the local Universe with a reasonable degree of accuracy. As we
describe in detail in the next section, GALICS has been
specifically tested against a wide variety of multi-wavelength
(UV-optical) early-type galaxy properties across a wide range in
redshift ($0<z<1.5$), making it the ideal tool to map the
early-type progenitor set over the redshift range $0<z<1$.

We now turn briefly to the impact of the free parameters on our
subsequent analysis and explore the potential uncertainty in our
results, given the freedom that we may have in the values of the
parameters in the model. Recalling that the primary aim of the
paper is to map the photometric properties of the early-type
progenitor set, we note that the parameters that potentially
affect the analysis the most are (a) those that are responsible
for the morphological mix of the Universe (since this determines
the composition of the progenitor set at any given time) and (b)
those that strongly affect galaxy luminosities/colours (since the
probabilistic prescriptions designed to include late-type
progenitors are given as a function of luminosities and colours).

As described above the key parameter that drives morphology is the
merger law, in particular the transition mass ratio ($\sim1:3$) at
which morphological transformations takes place (i.e. disks are
disrupted and bulges form). However, the merger recipe is
constrained reasonably robustly through numerical simulations and
the galaxy population predicted by the model is consistent with
the morphological mix of observed Universe, indicating that the
composition of the progenitor set is predicted with an acceptable
level of accuracy across our target redshift range.

In a similar vein, galaxy luminosities/colours are driven by a
suite of parameters including the SF efficiency, SN feedback, IMF
and dust recipes. This set of parameters is constrained through
comparison of the fiducial model to the properties of observed
galaxies. While we have, in principle, freedom to change these
parameters, calibrations to multiple observational facts in the
real Universe, such as the ones described in the previous section,
severely reduce this freedom in practice. In other words, if
individual parameters are altered arbitrarily then the
reproduction of the galaxy properties in the model would fail. In
this sense, the analysis presented in the paper is stable, in that
they are driven by values of the free parameters that are based
either on observations/hydrodynamical simulations and which
reproduce the galaxy properties in the present-day Universe with a
reasonable degree of accuracy.}

\begin{table*}
{\color{black}
\begin{center} \caption{Free parameters that affect
the evolution of the baryonic Universe, see Section 2.3 for more
details.}

\begin{tabular}{cccc}
\small

    Parameter   & Description                             & Fiducial value & Source of constraints\\ \hline \hline

    $\beta$     & Inverse of star formation efficiency    & 50             & Kennicutt (1998)\\
    $\epsilon$  & Inverse of mass loading for feedback    & 0.1            & \citet{Martin2002}\\
    $\chi$      & Galaxy merger power law                 & 3.333          & Numerical simulations e.g. Walker et al. (1996)\\\\\hline\\
    $\Omega_B$  & Baryon fraction                         & 0.02 $h^{-2}$  & $^2$H abundance in QSO absorption lines \citep{Tytler1996}\\
    $\psi$      & S-S merging normalisation               & 0.017          & \citet{Makino1997}\\
    $\varsigma$ & Recycling efficiency                    & 0.3            & Set to fiducial value in Hatton et al. (2003)\\
    $\kappa$    & Ratio of burst-to-bulge radius          & 0.1            & Set to fiducial value in Hatton et al. (2003)\\
    Dust        & ISM Extinction and emission due to dust & -              & Guiderdoni et al. (1987), Desert et al. (1990)\\
    IMF         & Initial mass function                   & -              & Kennicutt (1983)\\

\normalsize
\end{tabular}
\end{center}
\label{tab:constraint_comparisons}}
\end{table*}


\subsection{Reproduction of the early-type galaxy population}

{\color{black} Since our aim is to map the photometric properties
of early-type progenitors, it is important that the model
reproduces the multi-wavelength properties of the early-type
population and their evolution with redshift. A well-calibrated
model is clearly needed for the progenitor predictions to be
reliable.{\color{black} Note that in the redshift interval ($z<1$)
and for the wavelength range ($B$ and $V$ band) studied in this
paper, the population synthesis models employed by the fiducial
GALICS model - STARDUST \citep{Devriendt1999} - provide virtually
identical results to other commonly used models
\citep{Yi2003,BC03,Maraston2005}.}

{\color{black}In addition to the reproduction of the general
galaxy population, the fiducial GALICS model has been specifically
tested in the context of early-type galaxies, across virtually the
entire redshift range over which early-types have been observed
($0<z<1.5$).} GALICS accurately reproduces the optical colour
magnitude relations (CMRs) of the early-type population (in dense
environments) and their evolution from $z=0$ to $z \sim 1.23$
\citep{Kaviraj2005a}. The predicted evolution of both the gradient
and the scatter in the optical CMRs are consistent within errors
with various observational studies which use a variety of optical
colours (see Figure 8 in \citet{Kaviraj2005a}). More massive
ellipticals are predicted to be older (although they assemble more
recently) and more metal-rich than their less massive counterparts
(see Figure 3 in \citet{Kaviraj2005a}). The star formation
histories of elliptical galaxies are shown to be quasi-monolithic,
which enables the model to maintain the correct gradient and
scatter over the entire range in redshift ($0<z<1.27$) at which
observational early-type studies have been conducted (see Figure 5
in \citet{Kaviraj2005a}). Note that the decoupling of the assembly
history of elliptical galaxies from their star formation history
described above has also been found in other semi-analytical work
\citep[e.g.][]{Kauffmann1996,Baugh1996,deLucia2006}, although they
do not study the evolution of the optical colours with redshift.
Finally, GALICS is the only semi-analytic model to have been
tested against the new generation of $UV$ photometric data,
shortward of $3000 \AA$, made available by the recent GALEX
mission. Kaviraj et al. (2007) find excellent quantitative
agreement between the predictions of this model and the observed
$UV$-optical photometry of $\sim2100$ early-type galaxies in the
nearby ($0<z<0.11$) Universe that have been imaged by both GALEX
and the SDSS DR3.

Given its good reproduction of early-type photometry across a wide
wavelength range and its successful prediction of early-type
colour evolution to high redshifts ($z\sim1.23$), GALICS is a
useful and well-calibrated tool with which to follow the
progenitor set of early-type galaxies and the evolution of that
progenitor set in the redshift range $0<z<1$. We note that the
emphasis in this study is not to focus on the properties (e.g.
star formation and assembly histories) of early-type remnants at
$z=0$ (which is the major thrust of previous semi-analytical
studies of elliptical galaxies), but to focus on the properties of
the progenitor galaxies over the redshift range covered by the
bulk of the recent and forthcoming observational surveys
($0<z<1$).}}

\section{Dissecting the progenitor set: morphologies of progenitors and epochs of last mergers}
{\color{black}Early-type galaxies have an assortment of star
formation histories (SFHs). Central to this study is the epoch at
which the last merger, that finally creates the early-type
remnant, takes place. After this event, the early-type remnant
evolves to present day without further interactions with other
galaxies. In the discussion that follows, we refer to the
look-back time to this `last-merger' as the \emph{dynamical age}
of a galaxy. Note that the last mergers typically have mass ratios
of 1:5 or higher.}

As our subsequent analysis of the progenitor population involves
environments of model galaxies, an explanation of the definition
of this quantity is necessary. Galaxy environments in the model
are driven by the mass of the dark matter (DM) halo in which they
are embedded. At $z=0$, DM halo masses greater than $\sim 10^{14}
M_{\odot}$ correspond to `cluster' environments, while halo masses
between $\sim 10^{13} M_{\odot}$ and $\sim 10^{14} M_{\odot}$
correspond to `groups'. All other halo masses correspond to the
`field'. At higher redshifts these definitions do not strictly
hold since the DM halo population is evolving - for example, the
largest haloes at $z=1$ are likely to be roughly half their size
at present day \citep[e.g][]{VandenBosch2002}. We take this mass
accretion history into account when specifying the environments of
galaxies at high redshift. The mass accretion history is taken
from \citet[][see their Figure 5]{VandenBosch2002}.

{\color{black}The number of early-types galaxies in the
simulation, based on the definitions presented above, is 5418. 279
early-types are in clusters ($\sim 5\%$), 1081 ($\sim 20\%$) are
in groups and 4058 ($\sim 75\%$) are in the field. At $z=0$ there
are 8 clusters identified in the GALICS simulation box.}

\begin{figure}
\begin{center}
$\begin{array}{c}
\includegraphics[width=3.5in]{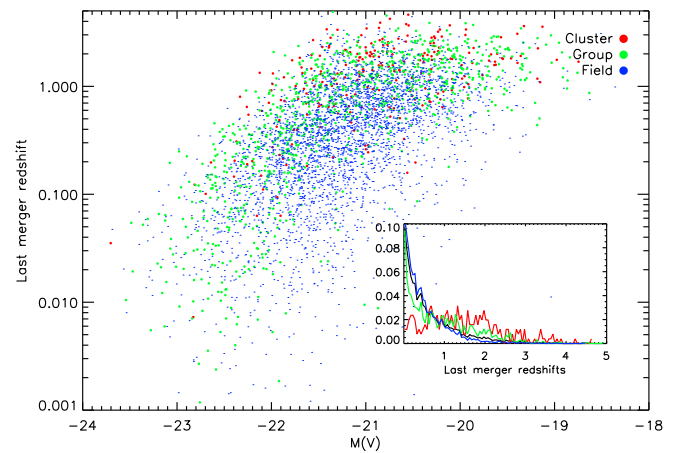}\\
\end{array}$
\caption{Last merger redshifts of the early-types in the model,
split by environment of the remnant at $z=0$. INSET: Histograms of
last merger redshifts shown in the top panel, split by
environment.} \label{fig:last_mergers}
\end{center}
\end{figure}

Figure \ref{fig:last_mergers} indicates the last merger redshifts
of the sample of early-types in the model, split by the
environment of the remnant at $z=0$. {\color{black} Recall that
after the last merger, the early-type remnant evolves to present
day without further interactions with other galaxies.} The inset
in Figure \ref{fig:last_mergers} shows histograms of the last
merger redshifts, again split by environment. As expected, we find
that, in all environments, larger early-types are assembled later
(although their stars are generally older \citep{Kaviraj2005a}.
This result is consistent with the findings of
\citet{deLucia2006}, although we note that they did not consider
early-types in different environments separately.

\begin{figure}
\begin{center}
\includegraphics[width=3.5in]{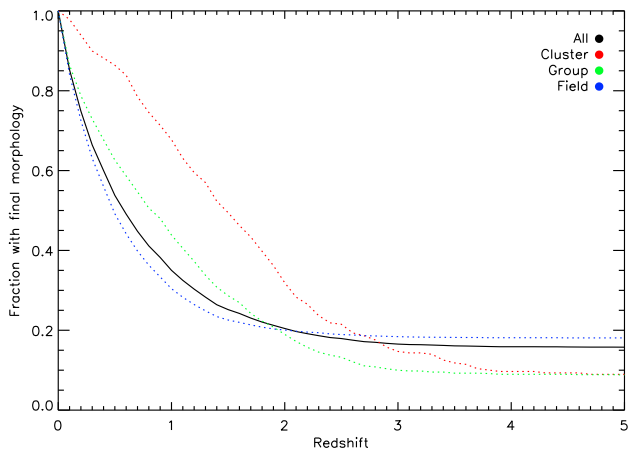}
\caption{Cumulative fraction of early-type galaxies which have had
their last merger as a function of redshift.}
\label{fig:final_morphs}
\end{center}
\end{figure}

Cluster galaxies (at least those brighter than $L_*$) have
significantly larger dynamical ages - morphological
transformations in clusters therefore proceed more quickly than in
all other environments. This point is made more clearly in Figure
\ref{fig:final_morphs}, where we plot the cumulative fraction of
early-type galaxies which have already had their last merger.
{\color{black}We find that, on average, without reference to
environment, only 35 percent of early-type galaxies are `in place'
(i.e. they evolve without further interactions with other
galaxies) by $z=1$ (black line). The rest are still `in pieces'.}
In terms of morphological transformations, cluster environments
are special, in that early-type morphologies are attained
significantly faster in clusters (red curve), with almost 70
percent of early-type galaxies having undergone their last merger
by $z=1$.

\begin{figure}
\begin{center}
\includegraphics[width=3.5in]{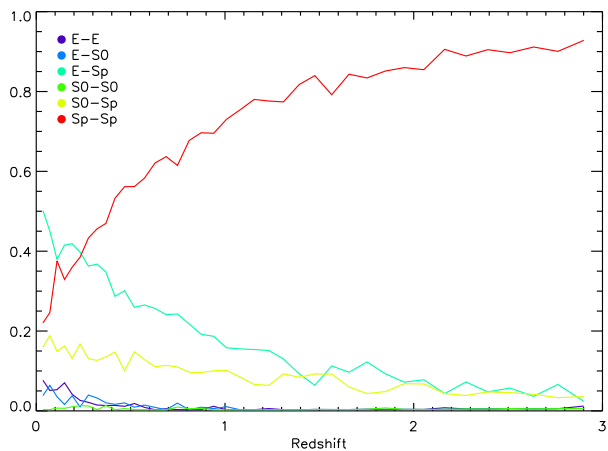}
\caption{Morphologies of progenitors in binary mergers as a
function of redshift. Non-binary mergers do happen but are rare,
and only take place, in the model, at redshifts greater than
$z\sim1.5$. Mergers at intermediate and high redshift are
dominated by pairs of progenitors which contain at least one
spiral progenitor.} \label{fig:merger_types}
\end{center}
\end{figure}

Before the last merger occurs, the morphology of the progenitors
is not necessarily early-type. Figure \ref{fig:merger_types} shows
the morphologies of progenitors in binarys mergers as a function
of redshift. Non-binary mergers do happen but are rare, and only
take place, in the model, at redshifts greater than 1. In
agreement with previous work \citep[e.g.][]{Khochfar2003} we find
that, in the local Universe, mergers between early-type
progenitors make up less than 20 percent of the merger activity.
All other mergers contain at least one spiral progenitor. Mergers
involving solely spiral progenitors increasingly dominate at
higher redshift and dominate the merger activity beyond $z=1$
\citep[see also][]{Kang2007}.

\begin{figure}
\begin{center}
\includegraphics[width=3.5in]{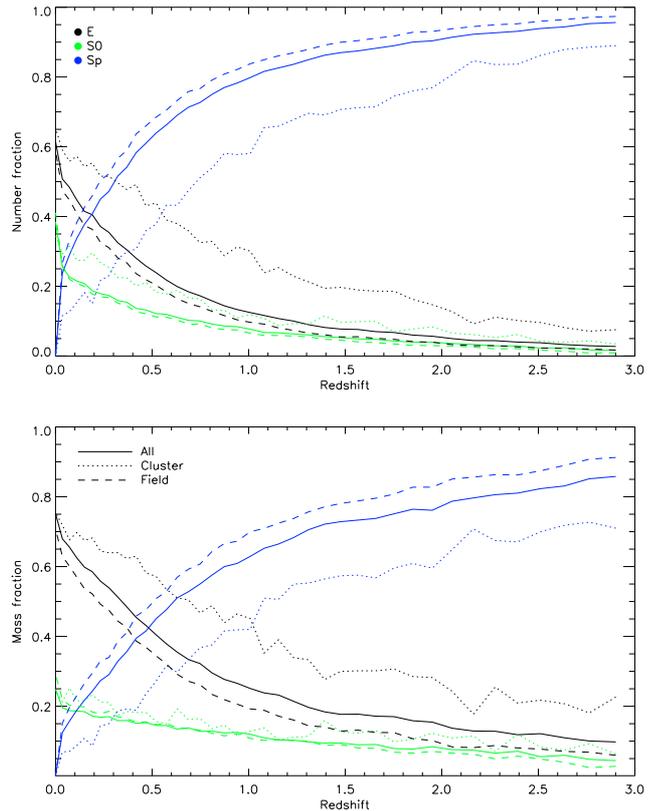}
\caption{Number (top) and mass (bottom) fractions contained in
progenitors of different morphological types in the redshift range
$0<z<3$, split by the environment of the early-type remnant.}
\label{fig:num_mass_env}
\end{center}
\end{figure}

\begin{figure}
\begin{center}
\includegraphics[width=3.5in]{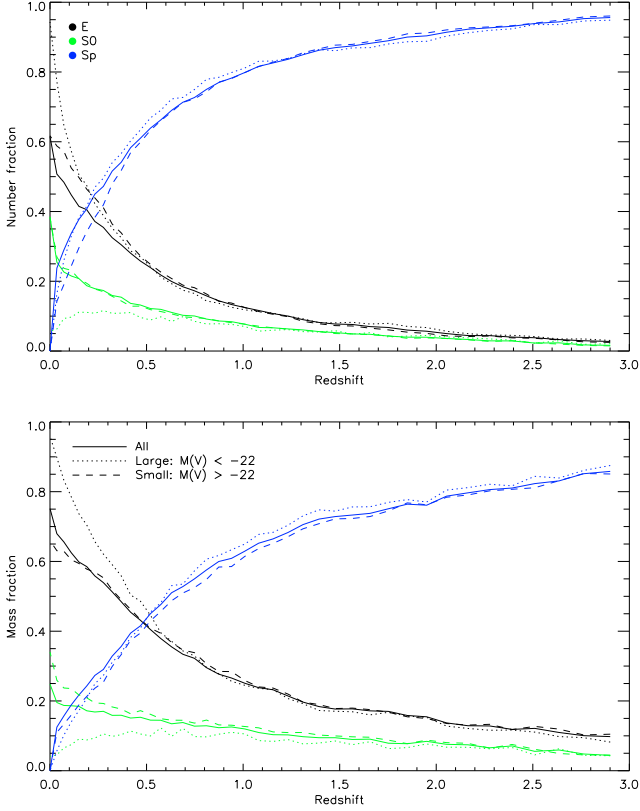}
\caption{Number (top) and mass (bottom) fractions contained in
progenitors of different morphological types in the redshift range
$0<z<3$, split by the luminosity of the early-type remnant at
present-day.} \label{fig:num_mass_lum}
\end{center}
\end{figure}

Having provided a picture of the merger activity within the
progenitor set, it is instructive to look at the fraction of the
progenitor set which is made up of a certain morphological type as
a function of redshift. Figures \ref{fig:num_mass_env} and
\ref{fig:num_mass_lum} show the number and mass fractions
respectively of progenitors of different morphological types, in
the redshift range $0<z<3$ and split by environment and luminosity
of the early-type remnant at $z=0$. We find that, averaging across
all environments, at $z\sim 1$, less than 50 percent of the
stellar mass which ends up in early-types today is actually in
early-type progenitors at this redshift. Faster morphological
transformations in cluster environments means that this value is
$\sim65$ percent in clusters at $z\sim1$. As a result, looking
only at early-type galaxies at $z\sim1$ does not take into account
almost half the stellar mass in the progenitor set. In other
words, the mass in the progenitor set doubles between $z=1$ and
$z=0$. {\color{black}A similar observational result was found by
Bell at al. (2004) and Faber et al. (2007), who used the optical
`red sequence' as a proxy for the progenitor set of present-day
early-type galaxies (see also the discussion in Section 6 below)}.

\begin{figure}
\begin{center}
\includegraphics[width=3.5in]{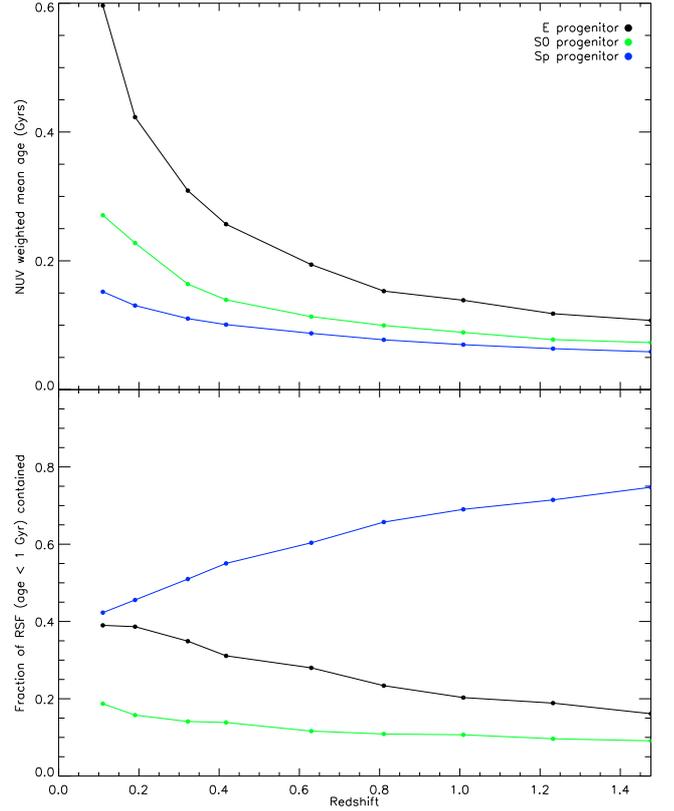}
\caption{TOP: Average $NUV$-weighted ages of progenitors of
different morphological types. The $NUV$ weighting is heavily
dominated by stars formed in these progenitors within the last 0.5
Gyrs of look-back time. The $NUV$ weighting was generated using
the GALEX (Martin el al. 2005) $NUV$ filter. BOTTOM: The mass
fraction of recently formed stars (age $<1$ Gyrs old) across the
progenitor set which is contained in progenitors of each
morphological type. Note that the bottom panel has to be
interpreted in conjunction with the mass fractions hosted by each
morphological type as a function of redshift. For example, at $z
\sim 0.1$, although spiral progenitors host $\sim$ 40 percent of
the total RSF in the progenitor set, they only constitute $\sim$
30 percent of the mass in the progenitor set (see bottom panel of
Figure \ref{fig:num_mass_env}). Early-type progenitors (elliptical
and S0 taken together) contribute $\sim60$ percent of the RSF -
but they also constitute $\sim 70$ percent of the total mass in
the progenitor set. Therefore at $z\sim0.1$ spiral progenitors
host 1.5 times the amount of RSF per unit mass than their
early-type counterparts.} \label{fig:rsf_contained}
\end{center}
\end{figure}

The bias does not arise simply because some progenitor mass is not
taken into account, but because the age profile of the mass in
progenitors of different morphological types tends to vary. We
illustrate this point in Figure \ref{fig:rsf_contained}. The top
panel shows the average $NUV$-weighted ages of progenitors of
different morphological types. The $NUV$ weighting, generated
using the GALEX (Martin el al. 2005) $NUV$ filter, is heavily
dominated by stars formed within the last 0.5 Gyrs of look-back
time. At all redshifts, early-type progenitors have higher
$NUV$-weighted ages, because the mass fraction contributed by
recent star formation (RSF) i.e. within the last 0.5 Gyrs is
smaller than for spiral progenitors. The differences between
elliptical and spiral progenitors are most pronounced at low
redshift. The bottom panel shows the fraction of the RSF across
the progenitor set that is contained in each morphological type.
This plot has to be interpreted in conjunction with the mass
fractions hosted by each morphological type as a function of
redshift. For example, at $z \sim 0.1$, although spiral
progenitors host $\sim$ 40 percent of the total RSF in the
progenitor set, they only constitute $\sim$ 30 percent of the mass
in the progenitor set (see bottom panel of Figure
\ref{fig:num_mass_env}). Early-type progenitors (elliptical and S0
taken together) contribute $\sim60$ percent of the RSF - but they
also constitute $\sim 70$ percent of the total mass in the
progenitor set. Therefore at $z\sim0.1$ spiral progenitors host
1.5 times the amount of RSF per unit masss than their early-type
counterparts. At higher redshift the balance of RSF contained in
each morphological type moves towards spiral progenitors, partly
because they are more spirals in the Universe than early-types.

Figure \ref{fig:rsf_contained} illustrates that an increasingly
larger fraction of RSF in the progenitor set is contained in
late-type systems at increasing redshift. In the context of
colour-magnitude relations (CMRs), which are often used to
age-date early-type populations at all redshifts, the exclusion of
spiral progenitors at high redshift biases the CMR towards redder
colours and does not give a proper indication of the age of all
the stellar mass that eventually constitutes present-day
early-type galaxies. This is particularly true if blue filters
(e.g. $U$-band or shorter wavelengths) are used in the age-dating
analysis.

\section{The spiral progenitors}
One of the aims of this study is to provide a means of including
spiral galaxies observed at high redshift which may be progenitors
into studies of early-type galaxy evolution, and thus correct, at
least partially, for progenitor bias. We therefore focus on spiral
progenitors predicted by the model and compare their photometric
properties to the general spiral population.

\subsection{The luminosity function of spiral progenitors}
\begin{figure*}
\begin{center}
\includegraphics[width=5in]{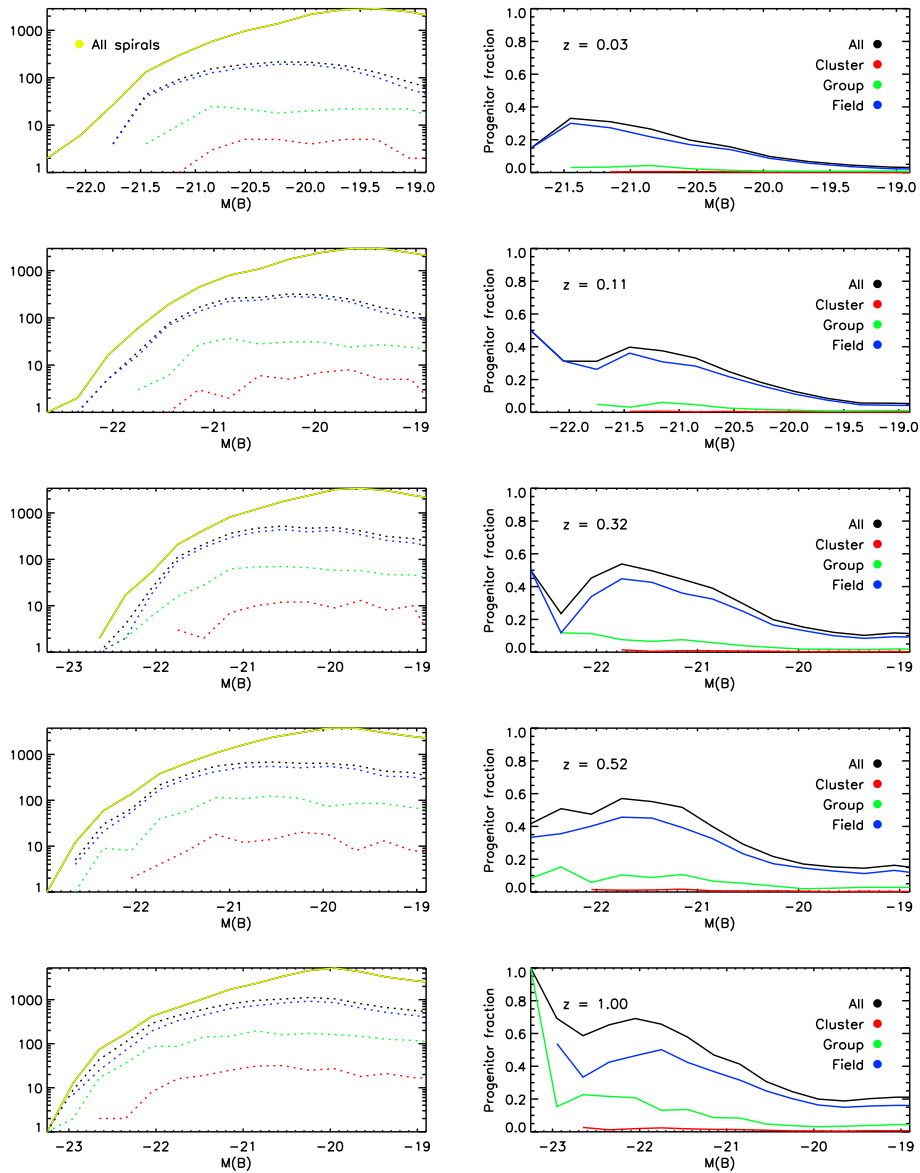}
\caption{Evolution of the $B$-band LF of spiral progenitors. The
left hand column illustrates the evolution of the spiral LFs - the
yellow curve denotes the LF of the general spiral population (i.e.
progenitors + non-progenitors) and the black curve the LF of
spiral progenitors only. The LFs of spiral progenitors whose
early-type remnants are, at $z=0$, in clusters, groups and the
field, are shown in red, green and blue respectively. The
right-hand column shows the fraction of spiral galaxies which are
early-type progenitors as a function of their B-band luminosities.
The colour coding is identical to that used for the left-hand
column.} \label{fig:b_spiralprogs}
\end{center}
\end{figure*}

We begin by studying the luminosity function (LF) of spiral
progenitors. We are interested in studying how the luminosities of
spiral progenitors compare to the general spiral population and
what fraction of spirals, at a given luminosity, are early-type
progenitors. In Figure \ref{fig:b_spiralprogs} we show the
evolution of the $B$-band LF of spiral progenitors - we also show
the spiral progenitors separated by the environment of their
corresponding early-type remnant at present-day. The left hand
column illustrates the evolution of the spiral LFs - the yellow
curve denotes the LF of the general spiral population (i.e.
progenitors + non-progenitors) and the black curve the LF of
spiral progenitors only. The LFs of spiral progenitors whose
early-type remnants are, at $z=0$, in clusters, groups and the
field, are shown in red, green and blue respectively. The
right-hand column shows the fraction of spiral galaxies which are
early-type progenitors as a function of their B-band luminosities.
The colour coding is identical to that used for the left-hand
column.



It is apparent that there is a greater preponderance of
progenitors amongst larger spirals at all redshifts. For example,
at low redshifts ($z<0.1$), 20 to 40 percent of spirals with
$M(B)<-20.5$ are early-type progenitors. At intermediate redshifts
($0.3<z<0.52$), these values rise to 30 and 60 percent
respectively. At high redshift ($z\sim1$) spirals with
$M(B)<-21.5$ have more than a 60 percent probability of being an
early-type progenitor, while spirals with $-20<M(B)<-21.5$ have
between a 30 and 40 percent chance of being early-type
progenitors. The falling progenitor fractions towards lower
redshift are partly due to the changing morphological mix of the
Universe.

\begin{figure*}
\begin{center}
\begin{minipage}{126mm}
$\begin{array}{c}
\includegraphics[width=5in]{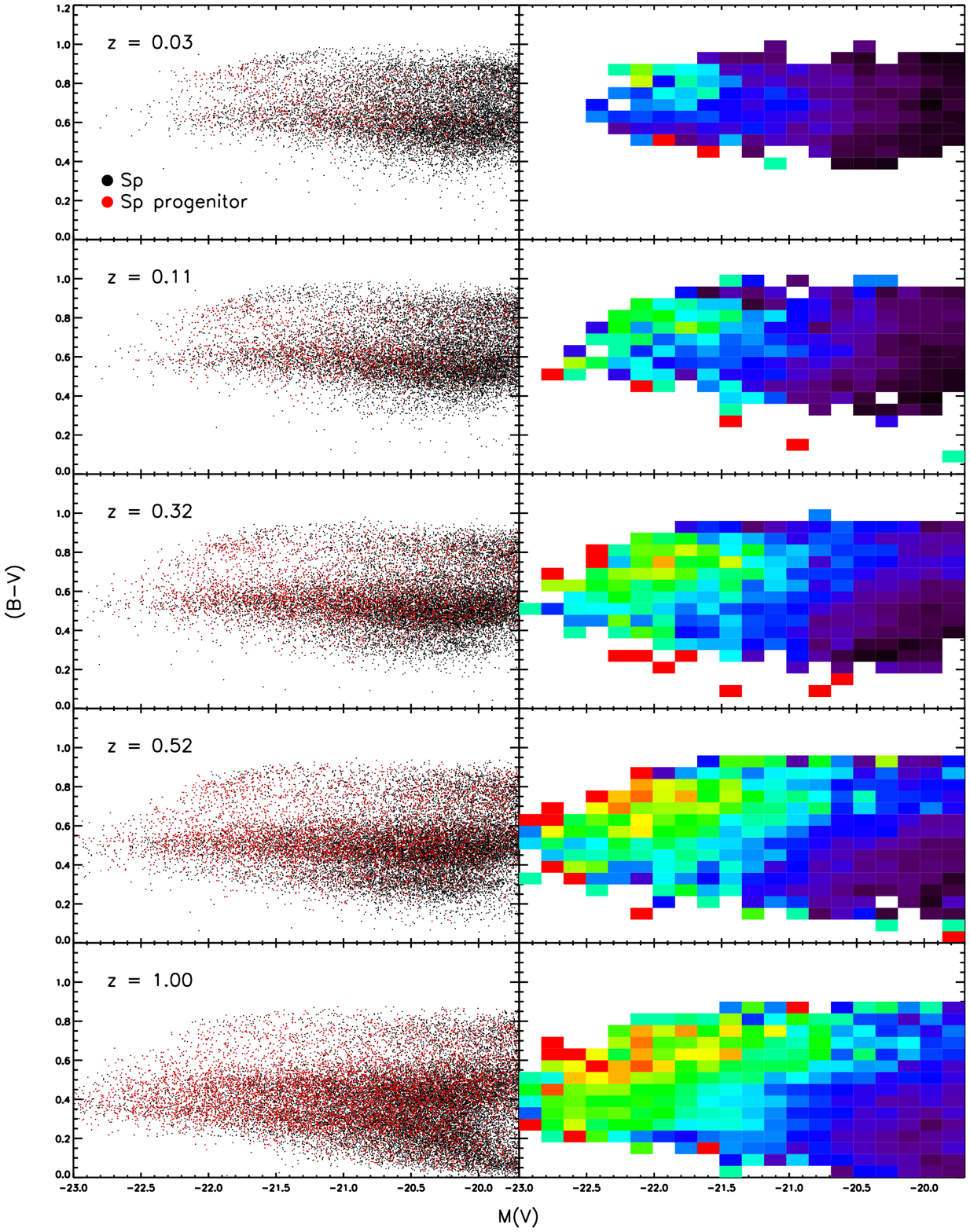}
\includegraphics[width=0.5in]{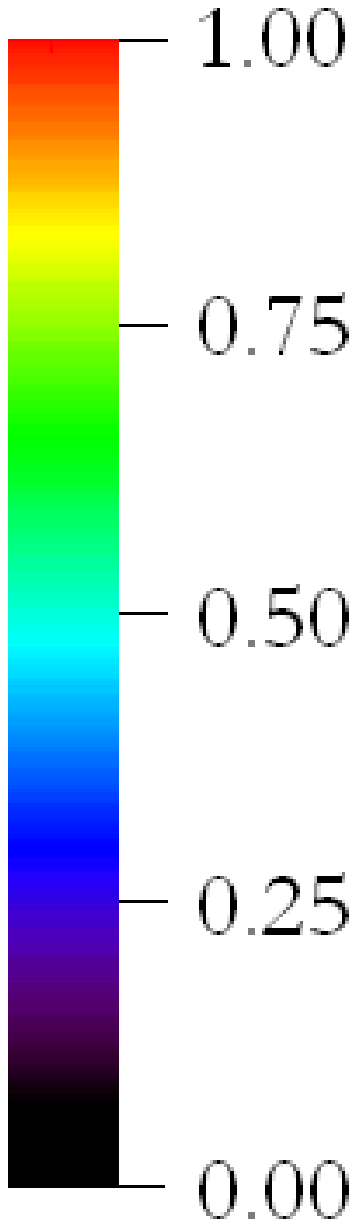}
\end{array}$
\caption{$(B-V)$ colours of the general population compared to the
$(B-V)$ colours of those spirals which are progenitors of
present-day early-type galaxies. Black dots represent spiral
galaxies and red dots represent spiral progenitors. The left-hand
column shows the spiral $(B-V)$ CMR from $z=0$ to $z=1$. In the
right hand column we show the corresponding (binned) fraction of
spiral progenitors across the $(B-V)$ CM space. The fraction
values are indicated using the colour coding shown in the legend.
Warmer colours indicate a higher progenitor fraction (red implies
a progenitor fraction of 1, black represents a progenitor fraction
of 0 and parts of the CM space without any galaxies are not
colour-coded).} \label{fig:bv_spiralprogs}
\end{minipage}
\end{center}
\end{figure*}


\subsection{The colour magnitude space of spiral progenitors}
While investigating the LFs of spiral progenitors is useful in
indicating the probability that a spiral of a given luminosity has
of being a progenitor, it is also desirable to explore the
colour-magnitude (CM) space of the spiral population, so that we
can separate progenitor spirals better from the general population
at a given luminosity.

In Figure \ref{fig:bv_spiralprogs} we compare the $(B-V)$ colours
of the general spiral population to the $(B-V)$ colours of spiral
progenitors. The left-hand column shows the spiral $(B-V)$ CMR
from $z=0$ to $z=1$. Black dots represent the spiral galaxies and
red dots represent spiral progenitors. In the right hand column we
show the fraction of spiral progenitors across the $(B-V)$ CMR.
The fractions are indicated using the colour coding shown in the
legend. Warmer colours indicate a higher progenitor fraction (red
implies a progenitor fraction of 1, black represents a progenitor
fraction of 0 and parts of the CM space without any galaxies are
not colour-coded).

At $z\sim0.1$, spirals with $-21.5<M(B)<-20.5$ have $\sim30$
percent chance of being a progenitor. For larger spirals, those
with red $(B-V)$ colours (i.e. $(B-V)>0.8$) have $\sim60$ percent
chance of being a progenitor, while the corresponding probability
for bluer spirals is 30 to 50 percent.

At intermediate redshift ($z\sim0.5$), black spirals, with
$-21.5<M(B)<-20.5$ and $(B-V)>0.6$, have $\sim30$ percent
probability of being an early-type progenitor, while blue spirals
in the same luminosity range have a low progenitor probability.
For larger spirals at these redshifts, the probabilities are
appreciably higher - red spirals with $(B-V)>0.7$ have between a
75 and 95 percent chance of being progenitors, while 50 to 75
percent of blue spirals in this luminosity range are progenitors.
The situation at high redshift $z\sim1$ is similar to that at
intermediate redshift.




\section{Progenitor evolution in clusters}
Before the advent of large scale surveys, dense regions of the
Universe were often targetted for early-type galaxy studies, both
at low and high redshift
\citep[e.g.][]{BLE92,Bower98,Stanford1998,VD1998,VD1999,VD2000,VD01}.
While studies of dense regions are attractive for a variety of
reasons \citep[e.g.][]{Ellis2002,VD2004}, a key benefit is
statistical convenience - clusters provide access to large
homogeneous samples of luminous objects at all redshifts. It has
been usual to `connect' results from cluster studies over large
redshift ranges to determine (at least qualitatively) the
chronology of galaxy evolution.

In this section, we investigate progenitors of present-day cluster
early-types, which are themselves in dense regions at $z>0$. The
motivation for this investigation is two fold. Firstly (and most
importantly), it provides a comparison to the vast literature of
`cluster' early-type studies. Secondly, this version of the GALICS
model has been accurately calibrated to match the optical CMRs of
early-types in dense regions from low to high redshift
\citep{Kaviraj2005a}, which implies that the colours of early-type
progenitors (regardless of morphology) are robustly reproduced.

It is important here to clarify the definitions of `density' that
we use to define both `clusters' at present-day, and `dense'
regions at high redshift. As mentioned before, model `density' is
assumed to be a direct function of the mass of the DM halo in
which galaxies are embedded. At $z=0$, a DM halo mass of
$10^{14}M_\odot$ represents the lower limit for a cluster-hosting
halo. Observational studies of dense regions at high redshift are
likely to contain an assortment of cluster-type haloes of varying
occupancies. Furthermore, DM haloes themselves are evolving - on
average, the largest haloes at $z=1$ are likely to be roughly half
their size at present day \citep[e.g][]{VandenBosch2002}. To take
these two points into account, we use a variable lower mass limit
for `cluster-hosting' haloes at $z>0$. At a given redshift this
lower limit is calculated from the average mass accretion history
\citep[][see their Figure 5]{VandenBosch2002} applied to a
$10^{14}M_\odot$ halo.

\begin{figure}
\begin{center}
\includegraphics[width=3.5in]{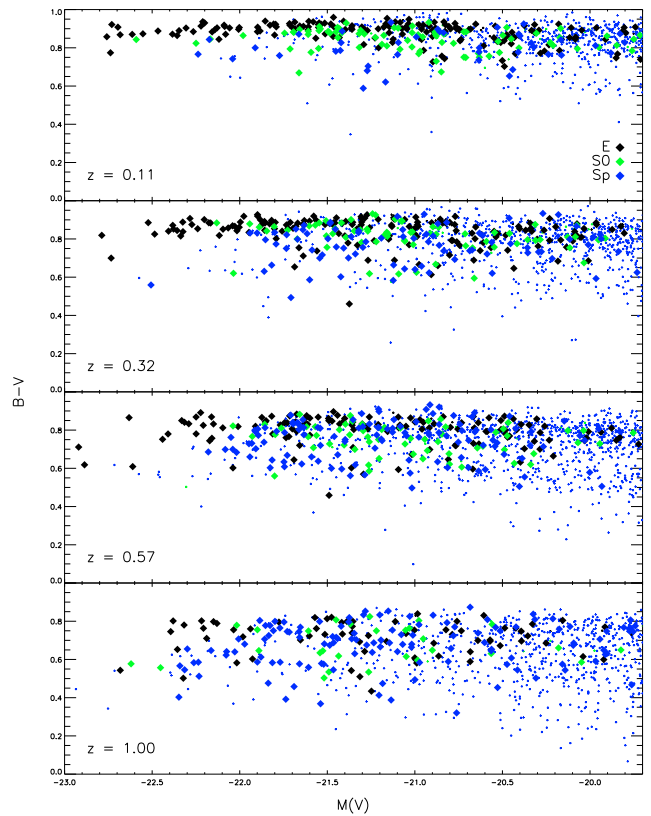}
\caption{The $(B-V)$ CMR in clusters in the redshift range
$0<z<1$. Large diamonds indicate the progenitors of present-day
cluster early-type galaxies. Small crosses indicate galaxies which
do not contribute to the mass in present-day cluster early-types.
Note that all elliptical and S0 galaxies in dense regions are, not
unexpectedly, progenitors of present-day cluster early-types.}
\label{fig:cluster_cmr}
\end{center}
\end{figure}

\begin{figure}
\begin{center}
\includegraphics[width=3.5in]{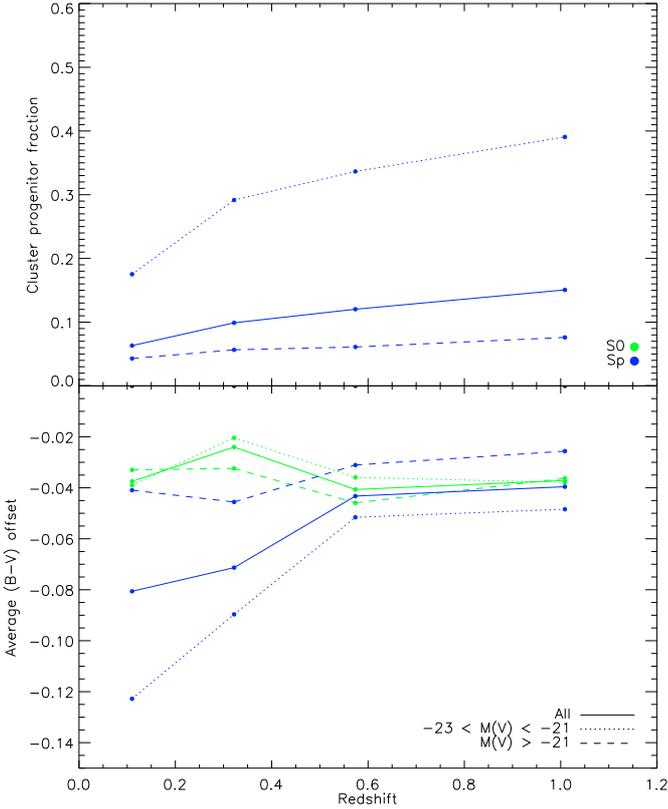}
\caption{TOP PANEL: The fraction of spiral galaxies in dense
regions (split by luminosity) which are progenitors of cluster
early-types at $z=0$. {\color{black}Note that all elliptical and
S0 galaxies in dense regions are, not unexpectedly, progenitors of
present-day cluster early-types}. BOTTOM PANEL: The offset in
average $(B-V)$ colour, with respect to elliptical progenitors, of
S0 and spiral progenitors in dense regions. The offsets are shown
split by luminosity.} \label{fig:clusterprog_props}
\end{center}
\end{figure}

Figure \ref{fig:cluster_cmr} shows the $(B-V)$ CMR in clusters in
the redshift range $0<z<1$. Large diamonds indicate progenitor
galaxies - black indicates ellipticals, green indicates S0s and
blue indicates spiral galaxies. Small crosses indicate galaxies
which do not contribute to the mass in present-day cluster
early-types. All early-type galaxies in dense regions are, not
unexpectedly, progenitors of cluster early-types at present day.
The top panel in Figure \ref{fig:clusterprog_props} shows the
fraction of spiral galaxies in dense regions at high redshift
(split by luminosity) which are progenitors of early-types at
$z=0$. The bottom panel shows the offset in $(B-V)$, with respect
to elliptical progenitors, of the S0 and spiral progenitor
galaxies. The offsets are shown split by luminosity.

We find that at high redshift ($z\sim1$), up to 40 percent of
large spirals ($-23<M(V)<-21$) are progenitors, whereas only
$\sim$10 percent of small spirals ($M(V)>-21$) are members of the
progenitor set. Large spirals are four times more likely to be
progenitors than small spirals, regardless of redshift, in the
redshift range $0<z<1$. Elliptical galaxies form the reddest locus
in $(B-V)$. S0 galaxies show an average offset of -0.04 compared
to the elliptical population, regardless of luminosity. Large
spiral progenitors show an average $(B-V)$ offset of -0.05
compared to the elliptical population at high redshift, mainly
because the scatter in the elliptical colours also tends to be
large at high redshift. At low redshift the offset is more
pronounced - large spiral progenitors are upto 0.1 mags bluer in
$(B-V)$ than the elliptical population.


\section{The `red sequence' as a proxy for the progenitor set}
\begin{figure}
\begin{center}
\includegraphics[width=3.5in]{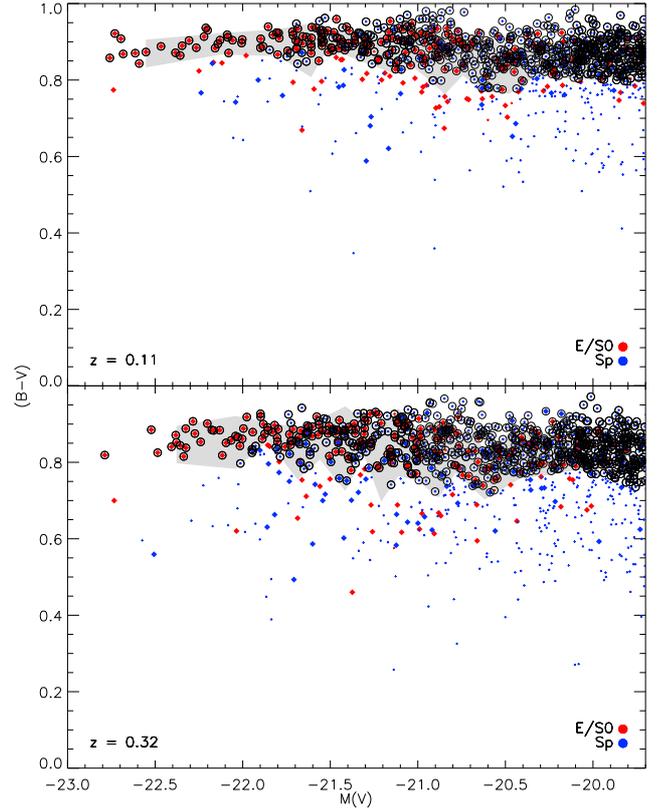}
\caption{The composition of the `red sequence', defined as the
galaxy population which occupies the part of the CM space
dominated by early-type galaxies (shown in grey), compared with
the progenitor set in cluster populations. The diamonds indicate
galaxies which are part of the progenitor set. Galaxies which are
not part of the progenitor set are shown using small crosses.
Galaxies in the red sequence are circled. The 'red sequence'
misses blue galaxies, both early-type and late-type. The top panel
shows the `red sequence' at $z=0.11$, while the bottom panel shows
the `red sequence' at $z=0.32$.} \label{fig:red_members1}
\end{center}
\end{figure}

\begin{figure}
\begin{center}
\includegraphics[width=3.5in]{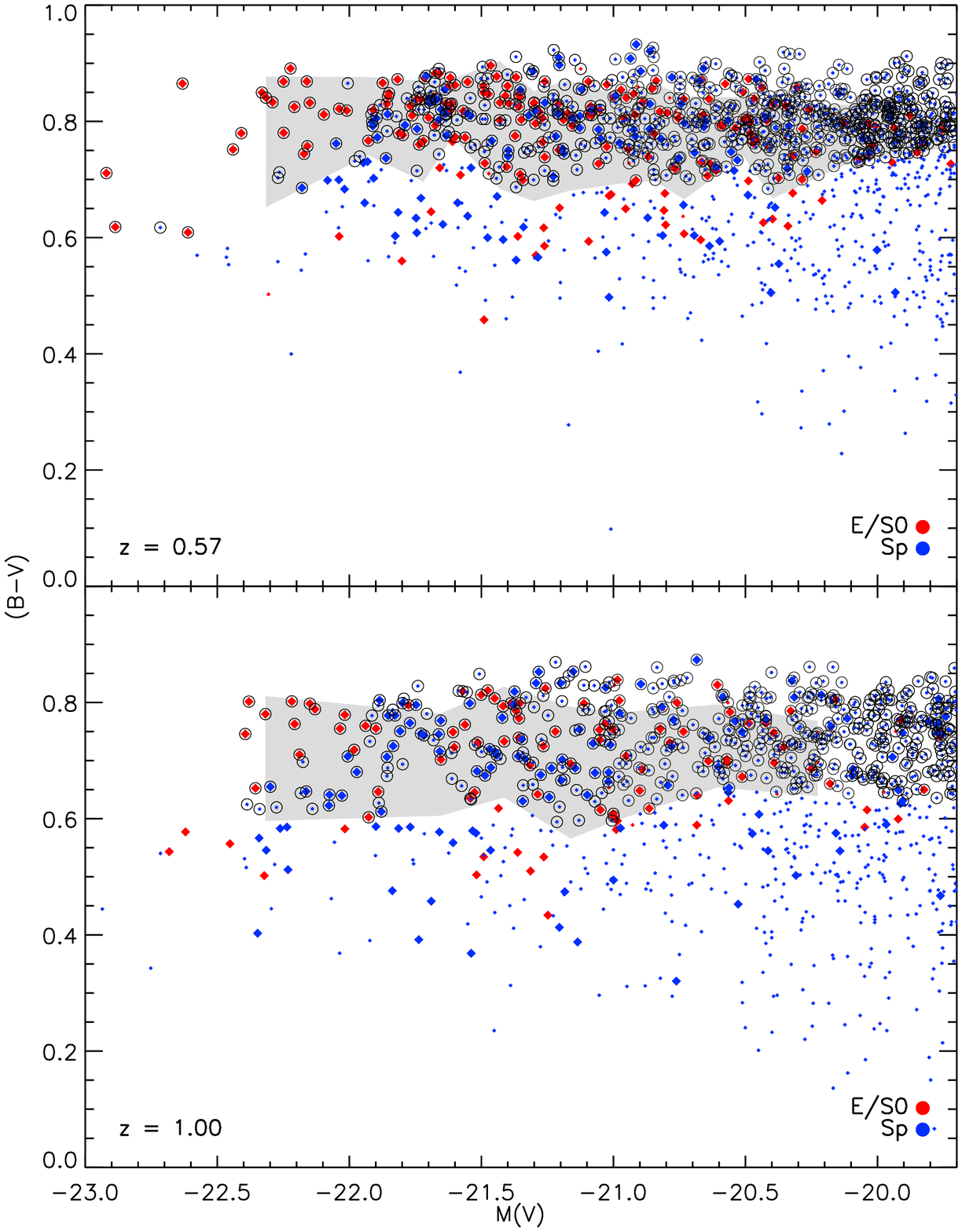}
\caption{The composition of the `red sequence', defined as the
galaxy population which occupies the part of the CM space
dominated by early-type galaxies (shown in grey), compared with
the progenitor set in cluster populations. The diamonds indicate
galaxies which are part of the progenitor set. Galaxies which are
not part of the progenitor set are shown using small crosses.
Galaxies in the red sequence are circled. The `red sequence'
misses blue galaxies, both early-type and late-type. This plot
shows the evolution of the `red sequence' from intermediate to
high redshift. The top panel shows the `red sequence' at $z=0.57$,
while the bottom panel shows the `red sequence' at $z=1$.}
\label{fig:red_members2}
\end{center}
\end{figure}

Early-type galaxies in clusters tend to preferentially populate
the reddest parts of the CM space. In this section we investigate
whether the sample of galaxies (without reference to morphology),
within the `red sequence' can be used as a proxy for the
progenitor set. We define the red sequence as the galaxy
population which occupies the part of the CM space which is
dominated by early-type galaxies. In Figures
\ref{fig:red_members1} and \ref{fig:red_members2} the CM space
dominated by early-type galaxies is shown in grey - this region is
determined by a progressive one-sigma fit to the colours of the
early-type population. It therefore contains, on average, 68
percent of the early-type population within it. Large diamonds
indicate galaxies which are part of the progenitor set. Galaxies
which are not part of the progenitor set are shown using small
dots. Galaxies in the red sequence are circled. It is apparent
that the red sequence misses blue progenitor galaxies, both
early-type and late-type. Figure \ref{fig:red_members1} shows the
evolution of the red sequence population from low to intermediate
redshift and Figure \ref{fig:red_members2} the corresponding
evolution from intermediate to high redshift.

\begin{figure}
\begin{center}
\includegraphics[width=3.5in]{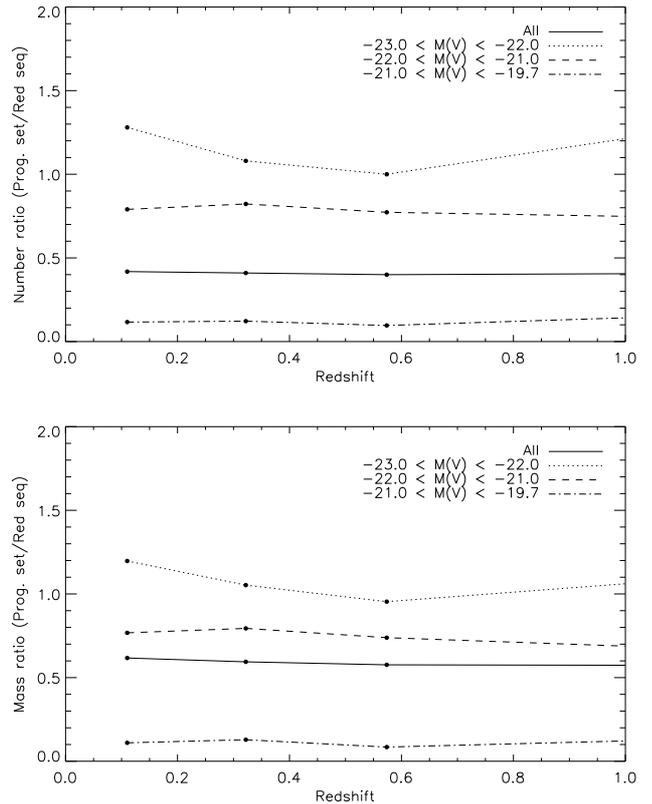}
\caption{Comparison between galaxies in the actual progenitor set
to those in the red sequence. TOP: Number ratio between the
progenitor set population and the red sequence population split by
redshift and luminosity. BOTTOM: Mass ratio between the progenitor
set population and the red sequence population split by redshift
and luminosity. It is apparent that large galaxies in the red
sequence trace the progenitor set well in terms of number and mass
but that the relationship breaks down rather rapidly as we move
towards the lower end of the luminosity function.}
\label{fig:mass_lum_ratio}
\end{center}
\end{figure}

In Figure \ref{fig:mass_lum_ratio} we compare galaxies in the
actual progenitor set to those in the red sequence. The top panel
shows the number ratio between the progenitor set population and
the red sequence population, split by redshift and luminosity,
while the bottom panel shows the mass ratio between the progenitor
set population and the red sequence population split by redshift
and luminosity. It is apparent that large galaxies
($-23<M(V)<-21$) in the red sequence trace the progenitor set well
in terms of number and mass but that the relationship breaks down
as we move towards the lower end of the luminosity function
($M(V)>-21$). The number and mass fractions remain stable, as a
function of the luminosity slices shown in Figure
\ref{fig:mass_lum_ratio}, within the redshift range explored in
this study ($0<z<1$).

{\color{black}Luminosity evolution studies which use the red
sequence as a proxy for the early-type population can therefore
achieve accurate results only for the massive end of the
luminosity function. Note that the results of two such recent
studies, Bell et al. (2004) and Faber et al. (2007), are accurate
since their analysis of the luminosity evolution of red galaxies
is indeed restricted to massive galaxies on the red sequence (e.g.
$M(V) \leq -21$ in Bell et al. 2004). However, it should be noted
that the red sequence is less reliable as a proxy for the
progenitor set further down the luminosity function. In addition,
the red sequence, almost by definition, misses contributions due
to early-types which lie blueward of it. This is increasingly true
at higher redshift - hence conclusions based on the colours of the
red sequence should be applied carefully to early-type evolution
as deeper or higher redshift data becomes available in the
future.}


\section{Summary and discussion}
{\color{black}We have presented a comprehensive theoretical study,
using a realistic semi-analytical model in the framework of the
standard LCDM paradigm, of the photometric properties of the
progenitors of present-day early-type galaxies. We have mapped, in
detail, the properties of these progenitors, regardless of their
morphology, in the redshift range $0<z<1$ (where the bulk of
current large-scale surveys are focussed), as a function of the
luminosity and environment of the early-type remnant at present
day. We have also developed probabilistic prescriptions which
provide a means of including spiral (i.e. non early-type)
progenitors at intermediate and high redshifts, based on their
luminosity and optical ($BVK$) colours. The prescriptions
developed here can potentially be used to address, from the
perspective of the standard modelss, the issue of `progenitor
bias', whereby the exclusion of late-type progenitors in
observational studies can lead to inaccurate conclusions regarding
the evolution of the early-type population over cosmic time. While
simple prescriptions designed to alleviate progenitor bias do
exist and have been used in some observational studies, they do
not reflect the mass assembly and morphological transformations
inherent to the standard model.

Against the backdrop of the impending or recent release of data
from large-scale surveys at high redshift, the results of this
study are hopefully timely, providing a picture of how the mass in
present-day early-types, across a range of luminosities and
environments, have assembled and exploring the extent to which
spiral progenitors play a role in this process. Our main
conclusions can be summarised as follows:}

\begin{itemize}

    \item Larger early-types in all environments are assembled
    later than their less massive counterparts. However, their
    stellar populations are generally older.
    {\color{black}This result is consistent with previous studies of early-type galaxies in the semi-analytical framework
    \citep[e.g.][]{Kauffmann1996,Baugh1996,Kaviraj2005a,deLucia2006}.}\\

    \item On average, without reference to environment, only 35
    percent of early-type galaxies are in place by $z=1$ {\color{black}and evolve without further interactions with other galaxies thereafter}.
    Morphological transformations are significantly faster in
    cluster environments (where the vast majority of early-type
    studies have been based before the advent of large-scale
    surveys). In clusters almost 70 percent of early-types are in
    place by $z=1$. In other words, the probability of a `major
    merger', which creates an early-type remnant is low after
    $z=1$ in cluster type environments.\\

    \item Averaging across all environments, at $z\sim 1$, less
    than 50 percent of the stellar mass which ends up in
    early-types today is actually in early-type progenitors at
    this redshift. This value is around 65 percent in clusters
    owing to faster morphological transformation in this
    environment. In other words, looking only at early-type
    progenitors does not take into account almost half the mass in
    the progenitor set - the progenitor set doubles in mass in the
    redshift range $0<z<1$.\\

    \item Progenitor bias does not arise simply because
    (late-type) progenitor mass is missed, but also because the age
    profile of mass in progenitors of different morphological
    types tend to vary. Spiral progenitors are typically `bluer'
    (at a given redshift) because they host more recently formed stars than early-type
    progenitors. Hence, age-dating the progenitor set using an
    `early-type only' CMR, i.e. after excluding the spiral
    progenitors, biases the CMR towards redder colours and overestimates the
    average age of the progenitor set.\\

    \item One of the aims of this study is to provide a
    means of including spiral progenitors into studies of early-type evolution, and thus correct, at least
    partially, for progenitor bias. We have therefore focussed on spiral
    progenitors in the model and compared their properties, in
    detail, to the general spiral population.\\

    \item There is a greater preponderance of progenitors amongst
    larger spirals at all redshifts. At
    low redshifts ($z<0.1$), 20 to 40 percent of spirals with
    $M(B)<-20.5$ are early-type progenitors. At intermediate redshifts
    $0.3<z<0.5$, these values rise to 30 and 60 percent respectively.
    At high redshift ($z \sim 1$) spirals with $M(B)<-21.5$ have more
    than a 60 percent probability of being an early-type progenitor
    while spirals with $-20<M(B)<-21.5$ have between a 30 and 40
    percent chance of being early-type progenitors. The falling
    progenitor fractions towards lower redshift are partly due to the
    changing morphological mix of the Universe.\\

    \item The colour-magnitude space of the spiral population
    provides a slightly better route to identifying spiral progenitors. At $z\sim0.1$,
    spirals with $-21.5<M(B)<-20.5$ have $\sim 30$ percent chance of
    being a progenitor. For larger spirals, those with red $(B-V)$
    colours, i.e. $(B-V)>0.8$, have $\sim60$ percent chance of being a
    progenitor, while the corresponding probability for bluer spirals
    is 30 to 50 percent.\\

    At intermediate redshift ($z\sim0.5$), black spirals, with
    $-21.5<M(B)<-20.5$ and $(B-V)>0.6$, have $\sim30$ percent
    probability of being an early-type progenitor, while blue spirals
    in the same luminosity range have a low progenitor probability.
    For larger spirals at these redshifts, the probabilities are
    appreciably higher - red spirals with $(B-V)>0.7$ have between a
    75 and 95 percent chance of being progenitors, while 50 to 75
    percent of blue spirals in this luminosity range are progenitors.
    The situation at high redshift $z\sim1$ is similar to that at
    intermediate redshift. The trends in the $(V-K)$ colour are similar to
    those in $(B-V)$.\\

    \item {\color{black}We suggest one example of how the probabilites
    derived in Section 4 could be used to `include' spiral progenitors at high redshifts
    ($z>0.5$). Modern optical surveys (e.g. MUSYC) can be used to trace
    the rest-frame $UV$ photometry of galaxies at high redshifts. Given its sensitivity to young
    stars, the rest-frame $UV$ is an excellent photometric tracer of recent star formation, i.e. the
    mass fraction of stars formed within the last Gyr. Clearly, if one
    had access to the rest-frame $UV$ over a range of redshifts, one could progressively
    trace the build-up of stellar mass
    in a particular class of galaxy over time. Thus, the stellar mass build-up in the early-type
    progenitor set can be approximated by the sum of the recent star
    formation in early-type galaxies within a given redshift range plus the recent star formation
    in spiral galaxies weighted by their relevant probabilities of
    being progenitors derived in Section 4.}\\

    \item {\color{black}We note that the analysis presented in this
    paper strongly implies that model and observed early-type populations should onlys be compared
    once they have been split by environment, since the assembly history of early-type galaxies is a
    function of environment. For example, the combined model population (at any redshift) should not be compared
    to an observed population which may be drawn predominantly from rich groups and
    clusters and vice-versa.}\\

    \item Finally we have explored the correspondence between the
    progenitor set and the `red sequence', defined as the part of
    the CM parameter space which is dominated by early-type galaxies.
    We find that galaxies, both late and early-type, that fall in
    this parameter space do not necessarily trace the progenitor
    set well. Large galaxies ($-23<M(V)<-21$) in the red sequence
    correspond to the progenitor set reasonably well in terms of number and mass but the
    relationship breaks down as we go towards the lower end of the luminosity function
    ($M(V)>-21$). Hence, luminosity evolution studies which use the red sequence as a proxy
    for the early-type population, therefore
    achieve accurate results only for the upper end of the luminosity
    function. In addition, the red sequence,
    almost by definition, misses contributions due to early-types which lie blueward
    of it - hence conclusions based on the colours of the red sequence should not
    generally be applied to early-type evolution, especially at
    high redshift.\\

\end{itemize}

\section*{Acknowledgements}
{\color{black}We are grateful to the anonymous referee for a
thoughtful review which added clarity to the original manuscript.
SK would like to thank J\'{e}r\'{e}my Blaziot for his generous
help with implementing GALICS and acknowledges a Research
Fellowship from the Royal Commission for the Exhibition of 1851, a
Senior Research Fellowship from Worcester College, Oxford and
support from the BIPAC institute at Oxford. Part of this work was
supported by a Leverhulme Early-Career Fellowship (SK, till Oct
2008). SKY was supported by the Korea Research Foundation Grant
funded by the Korean government (KRF-C00156). We thank Roger
Davies, Rachel Somerville, Richard Bower, Daniel Thomas, Claudia
Maraston, Christian Wolf, Kevin Schawinski and Sadegh Khochfar for
constructive comments on the manuscript and the work that led up
to it.}


\nocite{Blakeslee2003} \nocite{Kaviraj2007} \nocite{Martin2005}
\nocite{Bell2004} \nocite{Faber2007} \nocite{sdssdr4}
\nocite{Wolf2004} \nocite{Martin2005} \nocite{Gawiser2006}
\nocite{Rix2004} \nocite{Guiderdoni1987} \nocite{Desert1990}
\nocite{Kennicutt1983} \nocite{Loveday1996}


\bibliographystyle{aa}
\bibliography{references2}


\end{document}